\documentclass[12pt]{article}
\newcommand{\beq}{\begin{equation}}
\newcommand{\eeq}{\end{equation}}
\newcommand{\bra}{\begin{array}}
\newcommand{\era}{\end{array}}

\newcommand{\te}{\theta}
\newcommand{\al}{\alpha}
\newcommand{\ga}{\gamma}
\newcommand{\de}{\delta}

\newcommand{\Om}{\Omega}

\newcommand{\ep}{\epsilon}
\usepackage{graphicx}
\usepackage{latexsym}
\begin{document}
\author{Jamila Douari\footnote{jdouari@gmail.com}\\ \\
\small\it LTP, Hay Sa\^ada, Ain Taoujdate, Morocco\rm}
\title{Electrified Dp-Branes Intersections}
\maketitle
\frenchspacing
\pagenumbering{arabic}    

\section*{Abstract}
\hspace{.3in}In this review, we introduce the electrified Dp-branes intersections in the low energy effective theory. We focus on D1-D3, D1-D5 and D0-D2 branes. We give the solutions configurations in the low energy effective theory in the absence and the presence of electric field by exciting one, two and three scalars in D3 system. The solutions from D3 point of view in the last two cases are given as a spike which is interpreted as an attached bundle of a superposition of coordinates of another brane given as a collective coordinate a long which the brane extends away from the D3-brane. The lowest energy in both cases is higher than the energy found in the case of D1$\bot$D3 branes. We also find space-time dependent solutions in D1-D3 system that are a natural generalization of those found without the electric field. Then, we show that, in the flat background, the D1-D3 and D1-D5 branes obey Neumann boundary conditions by discussing the fluctuations of fuzzy funnel solutions in both systems. Also, we give the broken duality in D1-D3 system and the unbroken duality in D1-D5 system. In D0-D2 system, we consider generalized Maxwell theory by introducing a generalized connection put at the origin of the spherical D2-brane to describe anyons instead of Chern-Simons term. The D2-brane got a higher energy and the static potential for two opposite charged exotic particles is found to have a screening nature on a fuzzy two-sphere instead of confinement which is a special property of the system on the plane.
\vskip1cm PACS: 11.25.-w; 11.25.Uv; 11.27.qd; 11.15.Kc; 71.10.Pm
\vskip0.5cm Keywords: Intersecting Branes; Dyons, Duality; Boundary Conditions; Generalized connection; Exotic particles; Fuzzy two-sphere.
\newpage
\tableofcontents
\newpage
\section{Introduction}
\hspace{.3in}D-branes have brought about significant advances in string theory. They are known as extended objects on which string endpoints can live. Various brane configurations have attracted much attention over recent years and several papers have been devoted to the study of the relationship between noncommutative geometry \cite{ncg} and string theory \cite{ncst} and the relationship between D-branes with different dimensions as well \cite{brane,BI,InterBran1,InterBran2,dd3,cm,cm2,cm3,fuzzyQHE,f1}. The appearance of noncommutative geometry in string theory can be understood from a different point of view. For example, in type IIA theory, a D2-brane can be constructed from multiple D0-branes by imposing a noncommutative relation on their coordinates in matrix theory or under the strong magnetic field the world volume coordinates of a D2-brane become noncommutative by considering the quantum Hall system \cite{fuzzyQHE} and the magnetic field charge is interpreted as the number of D0-branes. On other hand, in type IIB theory, much of the progress has come about by directly studying the low energy dynamics of the D-branes world volume which is known to be governed by the Born-Infeld (BI) action \cite{BI,cm,NBI,dual} which has many fascinating features. Among these there is the possibility for D-branes to morph into other D-branes of different dimensions by exciting some of the scalar fields \cite{InterBran1,InterBran2}. The subject of intersecting branes in string theory is very rich and has been studied for a long time \cite{inter,fun}. The dynamics of their solutions (bion spike \cite{InterBran1,fun,dual,NBCBI} and fuzzy funnel \cite{fuzfun,dual}) were studied by considering linearized fluctuations around the static solutions \cite{fluct,d1d3}.

Among the goals of this review is to show that, in the presence of a world volume electric field, space-time dependent solutions can be generalised nicely and for certain condition we observe brane collapses with a speed less than that of light. We have also obtained a generalisation of the BPS solution. Another goal is to give the lowest energy in D3-brane theory by exciting two and three scalars \cite{dd3} and also to determine the boundary conditions of D1$\bot$D3 and D1$\bot$D5 branes in solving the equations of motion of the fuzzy funnel's fluctuations and discussing the associated potentials \cite{d1d3,d1d3d1d5}. We remark that the variation of the potential $V$ in terms of electric field $E$ and the spatial coordinate $\sigma$ in non-zero modes for both overall and relative transverse fluctuations in D1$\bot$D3 system and in zero mode of overall transverse fluctuations in D1$\bot$D5 system shows a singularity at some stage of $\sigma$. This is more clearly seen at the presence of electric field which leads to separate the system into two regions; small and large $\sigma$ depending on electric field. This implies that these intersecting branes obey Neumann boundary conditions and the end of open string can move freely on the brane. Consequently, the idea that the end of a string ending on a Dp-brane can be seen as an electrically charged particle is supported by this result. The obtained result in D1$\bot$D3 system is agree with its dual discussed in \cite{NBCBI} considering Born-Infeld action dealing with the fluctuation of the bion spike in D3$\bot$D1 branes case. The dual case of D1$\bot$D5 branes is not yet discussed.

Another interesting subject we discuss, in this work, concerns the duality of D1$\bot$D5 branes in the presence of electric field \cite{dd3}. By considering the energy of our system from D1 and D5 branes descriptions we find that the energies obtained from the two theories match and the presence of an electric field doesn't spoil this duality. By contrary, in the case of D1$\bot$D3 branes showed in \cite{brokdual}, the duality is broken because of the presence of electric field.

Then, another important point we deal with in this review concerns the spherical D2-brane \cite{cm3}. We consider exotic particles described by generalized Maxwell theory in which we introduce a generalized connection on a fuzzy two-sphere which has a dual description in terms of an abelian gauge field on a spherical D2-brane and is interpreted as a bound state of a spherical D2-brane and D0-branes. The exotic particles are known as excitations and quasi-particles or anyons; i.e. fermions (bosons) carrying odd (even) number of elementary magnetic flux quanta \cite{quanta}. They are living in two-dimensional space as composite particles having arbitrary spin, and they are characterized by fractional statistics which are interpolating between bosonic statistics and fermionic one \cite{quanta,anyon}. Among the main results in this section is the change of the potential's nature; there is no confinement any more, and the disappearance of the confinement in the two-sphere case for the exotic system is very interesting result. We also find that the energy of the gauge field dominates when the radius of the fuzzy two-sphere goes to infinity, then the energy of flat D2-brane which is a dual of the fuzzy two-sphere becomes high which is different from the case of the quantum Hall effect (QHE) where the energy of the flat D2-brane goes to zero. An important remark is that our system could be identified with the QHE in higher dimensions only if the radius $r$ and the number of charges $N$ go to zero. Also, what makes the model very different and very special is that the potential loses the confining nature in the fuzzy two-sphere case.

This review is organized as follows. We begin in section 2 with a brief review on D1$\bot$D3 and D1$\bot$D5 branes in dyonic case by using the abelian and non-abelian BI actions. In section 3, we investigate the two and three excited scalars in D3-brane theory. In section 4, we give the generalization space-time dependent solutions in the presence of a world volume electric field. In section 5, we study the electrified fluctuations of fuzzy funnel solutions corresponding to D3 and D5 branes. In the first subsection, we review the zero and non-zero modes of the overall transverse electrified fluctuations of fuzzy funnel solutions given in \cite{d1d3}. Then we discuss the zero and non-zero modes of the relative transverse fluctuations in the second subsection. In the third subsection, we treat the electrified fluctuations of the fuzzy funnel in D1$\bot$D5 system. We study the solutions of the linearized equations of motion of the overall transverse fluctuations in zero mode and we discuss its associated potential at the extremities of $\sigma$. In section 6, we review in brief the broken duality of D1$\bot$D3 in the dyonic case and the unbroken duality of intersecting D1-D5 branes in the presence of an electric field. The section 7, the fuzzy two-sphere is realized as one of D2-brane descriptions with special properties because of the generalized Maxwell theory. The conclusion is presented in section 8.

\section{Intersecting Dp-Branes}
\hspace{.3in}In this section, we review in brief the intersection of D1 branes with D3 and D5 branes. We focus our study on the presence of electric field and its influence on the potentials and the fluctuations of the fuzzy funnels.
\subsection{Electrified D1$\bot$D3 System}
\hspace{.3in}We start by giving in brief the known solutions of intersecting D1-D3 branes. From the point of view of D3 brane description the configuration is described by a monopole on its world volume. We use the abelian BI action and one excited transverse scalar in dyonic case to give the bion solution \cite{cm,Gib}. The system is described by the
following action \beq\bra{lll} S=\int dt L &=-T_3 \int
d^4\sigma\sqrt{-det(\eta_{ab}+\lambda^2\partial_a \phi^i \partial_b \phi^i +\lambda F_{ab})}\\\\
&= -T_3 \int d^4\sigma\Big[ 1 +\lambda^2 \Big( \mid \nabla\phi \mid^2 +\stackrel{\rightarrow}{B}^2 +\stackrel{\rightarrow}{E}^2 \Big)\\\\
&+\lambda^4 \Big( (\stackrel{\rightarrow}{B}.\nabla\phi )^2 +(\stackrel{\rightarrow}{E}.\stackrel{\rightarrow}{B})^2 +\mid \stackrel{\rightarrow}{E}\wedge\nabla\phi\mid^2 \Big) \Big]^{\frac{1}{2}} \era\eeq in which
$F_{ab}$ ($a,b=0,...,3$) is the field strength and the electric field is denoted as
$F_{0a}=E_a$. $\sigma^a$ denote the world volume coordinates while $\phi^i$
($i=4,...,9$) are the scalars describing transverse fluctuations of
the brane and $\lambda=2\pi \ell_s^2$ with $\ell_s$ is the string
length. In our case we excite just one scalar so $\phi^i=\phi^9
\equiv\phi$. Following the same process used in the reference
\cite{cm} by considering static gauge, we look for the lowest energy
of the system. Accordingly to (1) the energy of dyonic system is
given as \beq\bra{ll} \Xi&= T_3 \int d^3\sigma\Big[ \lambda^2 \mid
\nabla\phi +\stackrel{\rightarrow}{B}
+\stackrel{\rightarrow}{E}\mid^2 +(1-\lambda^2
\nabla\phi.\stackrel{\rightarrow}{B})^2-2\lambda^2
\stackrel{\rightarrow}{E}.(\stackrel{\rightarrow}{B}
+\nabla\phi)\\\\
&+ \lambda^4 \Big(
(\stackrel{\rightarrow}{E}.\stackrel{\rightarrow}{B})^2 +\mid
\stackrel{\rightarrow}{E} \wedge\nabla\phi\mid^2\Big)
\Big]^{1/2}.\era\eeq Then if we require $\nabla\phi
+\stackrel{\rightarrow}{B} +\stackrel{\rightarrow}{E}=0$, $\Xi$
reduces to \beq\bra{ll} \Xi_0 &=T_3 \int
d^3\sigma\Big[(1-\lambda^2
(\nabla\phi).\stackrel{\rightarrow}{B})^2+2\lambda^2
\stackrel{\rightarrow}{E}.\stackrel{\rightarrow}{E} )\\\\&+
\lambda^4 ( (\stackrel{\rightarrow}{E}.\stackrel{\rightarrow}{B})^2
+{\mid\stackrel{\rightarrow}{E}}\Lambda \nabla\phi\mid^2)
\Big]^{1/2}\era\eeq as minimum energy. By using the Bianchi identity
$ \nabla.\stackrel{\rightarrow}{B}= 0$ and the fact that the gauge field is static, the bion solution is then \beq\phi =\frac{N_m +N_e}{2r},\eeq
with $N_m$ is magnetic charge and $N_e$ electric charge.

Now we consider the dual description of the D1$\bot$D3 branes from D1
branes point of view. To get D3-branes from D-strings, we use the
non-abelian BI action. The natural definition of this action suggested in \cite{STr} is based on replacing the field strength in the BI action by a non-abelian field strength and adding the symmetrized trace $STr(\dots)$ in front of the $\sqrt{det}$ action. The precise prescription proposed in \cite{STr} was that inside the trace one takes a symmetrized average over orderings of the field strength. We refer the reader to \cite{STr} for more details on this action.

The non-abelian BI action describing D-string opening up into a D3-brane is given by
\beq S=-T_1\int d^2\sigma STr \Big[
-det(\eta_{ab}+\lambda^2 \partial_a \phi^i Q_{ij}^{-1}\partial_b
\phi^j )det Q^{ij}\Big]^{1\over2}\eeq where $Q_{ij}=\de_{ij}
+i\lambda \lbrack \phi_i , \phi_j \rbrack$.
Expanding this action to leading order in $\lambda$ yields the usual non-abelian scalar action
$$S\cong -T_1\int d^2\sigma  \Big[ N+ \lambda^2 Tr (\partial_a
\phi^i + \frac{1}{2}\lbrack \phi_i , \phi_j \rbrack \lbrack \phi_j ,
\phi_i \rbrack) +...\Big]^{1\over2}.$$ We deal with the leading order in $N$ when we expand the symmetrized trace and we consider large $N$ limit. The solutions of the equation of motion of the scalar fields $\phi_i$, $i=1,2,3$ represent the
D-string expanding into a D3-brane analogous to the bion solution of
the D3-brane theory \cite{InterBran1,InterBran2}. The solutions are
$$\bra{lc}\phi_i =\pm\frac{\al_i}{2\sigma},&\lbrack \al_i , \al_j
\rbrack=2i\ep^{ijk}\al_k ,\era$$ with the corresponding geometry is
a long funnel where the cross-section at fixed $\sigma$ has the
topology of a fuzzy two-sphere.

The dyonic case is presented by considering ($N, N_f$)-strings. We introduce a background U(1) electric
field on the $N$ D-strings, corresponding to $N_f$ fundamental strings dissolved on the world sheet \cite{dual}. The theory is described by the action \beq S=-T_1\int d^2\sigma STr \Big[
-det(\eta_{ab}+\lambda^2 \partial_a \phi^i Q_{ij}^{-1}\partial_b
\phi^j +\lambda F_{ab})det Q^{ij}\Big]^{1\over2}.\eeq The action can be rewritten as \beq S=-T_1\int d^2\sigma STr \Big[
-det\pmatrix{\eta_{ab}+\lambda F_{ab}& \lambda \partial_a \phi^j
\cr -\lambda \partial_b \phi^i & Q^{ij}\cr}\Big]^{1\over2}.\eeq By computing the determinant, the action becomes \beq
S=-T_1\int d^2\sigma STr \Big[ (1-\lambda^2 E^2 + \al_i \al_i
\hat{R}'^2)(1+4\lambda^2 \al_j \al_j \hat{R}^4 )\Big]^{1\over2},\eeq
in which we replaced the field strength $F_{\tau\sigma}$ by $EI_{N}$ ($I_{N}$ is
$N\times N$-matrix) and the following ansatz were inserted \beq\phi_i =\hat{R}\al_i
.\eeq Hence, we get the funnel solution for dyonic string by solving
the equation of variation of $\hat{R}$ as follows \beq \phi_i
=\frac{\al_i}{2\sigma\sqrt{1-\lambda^2 E^2}}.\eeq

\subsection{Electrified D1$\bot$D5 Branes}
\hspace{.3in}The fuzzy funnel configuration in which the D-strings expand into orthogonal D5-branes shares many common features with the D3-brane funnel. The action describing the static configurations involving five nontrivial scalars is
\beq \bra{ll} S&=-T_1 \int d^2\sigma STr\Big[ 1+\lambda^2
(\partial_\sigma \Phi_i)^2+ 2\lambda^2 \Phi_{ij}\Phi_{ji}+
2\lambda^4 (\Phi_{ij}\Phi_{ji})^2 -4\lambda^4
\Phi_{ij}\Phi_{jk}\Phi_{kl}\Phi_{li}\\\\&\phantom{~~}+2\lambda^4
(\partial_\sigma\Phi_{ij})^2 \Phi_{jk}\Phi_{kj}-4\lambda^4
\partial_\sigma \Phi_i \Phi_{ij}\Phi_{jk}\partial_\sigma \Phi_k
+\frac{\lambda^6 }{4}(\epsilon_{ijklm}\partial_\sigma \Phi_i
\Phi_{jk}\Phi_{lm})^2 \Big]^{1\over 2}, \era\eeq where
$\Phi_{ij}\equiv \frac{1}{2}\lbrack \Phi_i ,\Phi_j \rbrack$ and the
funnel solution is given by suggesting the following ansatz \beq \Phi_i
(\sigma)=\mp\hat{R}(\sigma)G_i,\eeq $i=1,...,5$, where
$\hat{R}(\sigma)$ is the (positive) radial profile and $G_i$ are the
matrices constructed in \cite{funnelSoluD5,f2}. We note that $G_i$ are
given by the totally symmetric $n$-fold tensor product of 4$\times$4
gamma matrices, and that the dimension of the matrices is related to
the integer $n$ by $N=\frac{(n+1)(n+2)(n+3)}{6}$. The Funnel
solution (12) has the following physical radius
\beq R(\sigma)=\frac{\lambda}{N}\sqrt{(Tr(\Phi_i)^2)}=\sqrt{c}\lambda\hat{R}(\sigma),\eeq with $c$ is the "Casimir" associated with the $G_i$ matrices, given by $c=n(n+4)$
and the resulting action for the radial profile $R(\sigma)$ is \beq
S=-NT_1 \int d^2\sigma  \sqrt{1+(R')^2}(1+4\frac{R^4}{c\lambda^2}).\eeq We note that this result only captures the leading large N contribution at each order in the expansion of the square root.

To extend the discussion to dyonic strings we consider ($N,N_f$)-strings. Thus, the electric field is on and the system dyonic is described by the action \beq S=-T_1\int d^2\sigma STr \Big[
-det(\eta_{ab}+\lambda^2 \partial_a \Phi^i Q_{ij}^{-1}\partial_b
\Phi^j +\lambda F_{ab})det Q^{ij}\Big]^{1\over2}.\eeq The action can be rewritten as
$$\phantom{~~~~~~~~~~~~~~~~~~~~~~~}S=-T_1\int d^2\sigma STr \Big[-det\pmatrix{\eta_{ab}+\lambda F_{ab}& \lambda \partial_a \Phi^j \cr -\lambda \partial_b \Phi^i & Q^{ij}\cr}\Big]^{1\over2},\phantom{~~~~~~~~~~~~~~~~~~}(15')$$
with $Q_{ij}=\de_{ij}+i\lambda \lbrack \Phi_i , \Phi_j \rbrack$ and $i,j=1,...,5$, $a,b=\tau,\sigma$. We insert the ansatz (12) and $F_{\tau\sigma}=EI_{N}$ ($I_{N}$ is $N\times N$-matrix) in the action (15). Then we compute the determinant and we obtain
$$\phantom{~~~~~~~~~~~~~~~~~~~~~~~~~~~~~}S=-NT_1 \int d^2\sigma\sqrt{1-\lambda^2 E^2+(R')^2}(1+4\frac{R^4}{c\lambda^2}).\phantom{~~~~~~~~~~~~~~~~~~}(15'')$$
The funnel solution is 
\beq \Phi_i (\sigma)=\mp\frac{R(\sigma)}{\lambda\sqrt{c}}G_i.\eeq
From $(15'')$ We can derive the lowest energy
$$\bra{ll} E&=NT_1 \int d\sigma\sqrt{\Big(\sqrt{1-\lambda^2 E^2}\pm R'\sqrt{\frac{8R^4}{c\lambda^2}+\frac{16R^8}{c^2\lambda^4}}\Big)^2 +\Big( R'\mp \sqrt{1-\lambda^2 E^2}\sqrt{\frac{8R^4}{c\lambda^2}+\frac{16R^8}{c^2\lambda^4}}\Big)^2}\\
&\ge NT_1 \int d\sigma\Big(\sqrt{1-\lambda^2 E^2}\pm R'\sqrt{\frac{8R^4}{c\lambda^2}+\frac{16R^8}{c^2\lambda^4}}\Big).\era$$ This is obtained when
$$R'=\mp \sqrt{1-\lambda^2 E^2}\sqrt{\frac{8R^4}{c\lambda^2}+\frac{16R^8}{c^2\lambda^4}}.$$ This equation can be explicitly solved in terms of elliptic functions. For small $R$, the $R^4$ term under the square root dominates, and we find the funnel solution. Then the physical radius of the fuzzy funnel solution (16) is found to be \beq R\approx\frac{\lambda\sqrt{c}}{2\sqrt{2}\sqrt{1-\lambda^2E^2}\sigma}.\eeq

In the dual point of view of the D5-brane world volume theory, the
action describing the system is the Born-Infeld action $$
S=-T_5\int d^6\sigma STr \sqrt{-det(G_{ab}+\lambda^2 \partial_a \phi
\partial_b \phi +\lambda F_{ab})},$$ with $a,b=0,1,...,5$ and
$\phi$ the excited transverse scalar. To get a spike solution with
electric field switched on from D5-brane theory we follow the
analogous method of bion spike in D3-brane theory as discussed in
\cite{cm,f1,f2}. We use spherical polar coordinates and the metric is
$$ ds^2 =G_{ab}d\sigma^a d\sigma^b =-dt^2 +dr^2 +r^2 g_{ij}d
\alpha^i d \alpha^j,$$ with $r$ the radius and $\alpha^i$,
$i=1,...,4$, Euler angles. $g_{ij}$ is the diagonal metric on a
four-sphere with unit radius $$g_{ij}=\pmatrix{1&&&\cr&sin^2
(\alpha^1)&&\cr&&sin^2 (\alpha^1)sin^2 (\alpha^2)&\cr&&&sin^2
(\alpha^1)sin^2 (\alpha^2)sin^2 (\alpha^3)\cr}.$$ In this theory,
we add the electric field $E$ as a static radial field in the $U(1)$
sector. The scalar $\phi$ is only a function of the radius by
considering bion spike solutions with a "nearly spherically
symmetric" ansatz. In the same time to compare with the radial
profile obtained in D1-brane theory we identify the physical
transverse distance as $\sigma=\lambda\phi$, and the radius $r =
R$ which fixes the coefficients. Thus the scalar is found to be $$
\phi(r)=\pm\int \frac{dr}{\lambda(1-\alpha^2)\Big(
(\frac{r^4}{\lambda^2 }+1)^2 -1\Big)},$$ where $\alpha= \frac{g_s N_f}{\sqrt{N^2 +g_s^2 N_f^2}}$.

\section{Exciting Two and Three Scalars}
\hspace{.3in}By following the same mechanism used for D1$\bot$D3 branes, this
section is devoted to finding the funnel solutions when two and three
excited scalars are involved. Thus, we restrict ourselves to using BPS
arguments to find some solutions which have the interpretation of a D2-brane ending on D3-brane
and D3-brane ending on other D3-brane. In these studies we consider
the absence and the presence of an electric field. We found that the
investigation of excited D3-brane in each case leads to
the fact that by exciting 2 and 3 of its transverse directions in
the absence or the presence of electric field, the brane develops a
spike which is interpreted as an attached bundle of a superposition
of coordinates of another brane given as a collective coordinate a
long which the brane extends away from the D3-brane. Then, by
considering the lowest energy states of our system we remark that the lowest
energy in the intersecting branes case is obtained by the
D1-D3 branes intersection and the energy is higher if we excite more
scalar fields and even more in the presence of an electric field.

Thus, we use the abelian Born-Infeld action for the world volume gauge field and transverse displacement scalars to explore some aspects of D3-brane structure and dynamics. We deal with magnetic and dyonic cases \cite{dd3}.
\subsection{Absence of Electric Field}
\hspace{.3in}We consider the case where D3-brane has more than one scalar
describing transverse fluctuations. We denote the world volume
coordinates by $\sigma^a$, $a=0,1,2,3$, and the transverse
directions by the scalars $\phi^i$, $i=4,...,9$. In D3-brane theory
construction, the low energy dynamics of a single D3-brane is
described by the BI action by using static gauge \beq S_{BI}=\int
L=-T_3 \int d^4\sigma\sqrt{-det(\eta_{ab}+\lambda^2\partial_a \phi^i
\partial_b \phi^i +\lambda F_{ab})} \eeq with $F_{ab}$ is the field
strength of the $U(1)$ gauge field on the brane. By
exciting two scalar fields and setting to zero the other scalars, the energy is evaluated for the fluctuations through two directions and for static configurations as follows

\beq\bra{ll} \zeta= -L&=T_3 \int d^3\sigma\Big[ 1 +\lambda^2 \Big(
\mid \nabla\phi_4 \mid^2 +\mid \nabla\phi_5 \mid^2
+\stackrel{\rightarrow}{B}^2 \Big)+ \lambda^4
\Big( (\stackrel{\rightarrow}{B}.\nabla\phi_4 )^2+(\stackrel{\rightarrow}{B}.\nabla\phi_5 )^2 \Big)\\\\
&+\lambda^4 \mid\nabla\phi_4 \wedge\nabla\phi_5 \mid^2 \Big]
^{\frac{1}{2}}. \era\eeq

If we introduce a complex scalar field $C=\phi_4 + i\phi_5 $, we
can rewrite the energy as \beq\bra{ll} \zeta&= T_3 \int
d^3\sigma\Big[ \lambda^2 (\nabla C+\stackrel{\rightarrow}{B})(\nabla
C+\stackrel{\rightarrow}{B})^{*}+(1-\lambda^2 (\nabla
C.\stackrel{\rightarrow}{B}) )(1-\lambda^2 (\nabla
C.\stackrel{\rightarrow}{B}))^{*}\\\\&+\frac{1}{4}\lambda^4 \mid
\nabla C \wedge\nabla C^{*} \mid^2\Big]^{\frac{1}{2}}. \era\eeq In
this case, we observe that to get minimum energy
we can set the first term to zero and this will lead to \beq\nabla C
=- \stackrel{\rightarrow}{B}=\nabla \phi_4 +i \nabla \phi_5 .\eeq We
know that $\stackrel{\rightarrow}{B}$ is real, then $\nabla \phi_5
=0$ and thus in the "bi-excited scalar" system the lowest energy is
identified to D3$\bot$D1 system. This suggests that to study
the minimum energy configuration of the D3-brane system it is only
worthwhile to excite just one scalar.

Now, requiring $\nabla \phi_5 \neq0$ we get different energy bound.
The energy (20) can be rewritten as follows \beq\bra{ll}
\zeta&=T_3 \int d^3\sigma\Big[ \lambda^2  \mid \nabla\phi_4
+\nabla\phi_5 \mp \stackrel{\rightarrow}{B} \mid^2 + (1\pm \lambda^2
\stackrel{\rightarrow}{B}.(\nabla\phi_4 +\nabla\phi_5))^2 \\\\
&\pm 2\lambda^2 \nabla\phi_4 .\nabla\phi_5 +\lambda^4 \mid
\nabla\phi_4 \wedge\nabla\phi_5 \mid^2\Big]^{\frac{1}{2}}. \era\eeq
The new bound is now found with the following constraint
\beq\nabla\phi_4+\nabla\phi_5 = \pm \stackrel{\rightarrow}{B},\eeq
and $\pm 2\lambda^2 \nabla\phi_4 .\nabla\phi_5 \geq 0$ should also be
satisfied. Then the energy is \beq \tilde{\zeta}= T_3 \int
d^3\sigma\Big[ (1\pm \lambda^2\stackrel{\rightarrow}{B}.(\nabla\phi_4 +\nabla\phi_5 ) )^2 +\lambda^4 \mid \nabla\phi_4 \wedge\nabla\phi_5 \mid^2\pm 2\lambda^2 \nabla\phi_4.\nabla\phi_5\Big]^{\frac{1}{2}}.\eeq We choose without loss of generality $\nabla\phi_4\perp\nabla\phi_5$.

Thus we get $\nabla^2 \phi_4 +\nabla^2 \phi_5 =0$ (by using the Bianchi identity), and the solution is
\beq
\phi_{4}+\phi_{5}=\pm\frac{N_m}{2r}.
\eeq

This solution could be generalized by considering three excited scalars. The energy is then found to be \beq\bra{ll}\zeta_3&=T_3\int d^3\sigma\Big[1+\lambda^2\Big(\sum\limits_{i=4}^{6}\mid \nabla\phi_i\mid^2 +\stackrel{\rightarrow}{B}^2\Big)+\lambda^4\Big(\sum\limits_{i=4}^{6}\mid\stackrel{\rightarrow}{B}.\nabla\phi_i\mid^2+\frac{1}{2}\sum\limits_{i,j=4}^{6}\mid \nabla \phi_i\wedge \nabla \phi_j\mid^2\Big) \Big]^{\frac{1}{2}}\\\\&\geq T_3\int d^3\sigma \Big[(1\pm\lambda^2\stackrel{\rightarrow}{B}.\sum\limits_{i=4}^{6}\nabla \phi_i)^2+\frac{1}{2}\sum\limits_{i,j=4}^{6}\mid\nabla\phi_i\wedge\nabla\phi_j\mid^2\pm\frac{\lambda^2}{2}\sum\limits_{i,j=4}^{6}\nabla\phi_i .\nabla\phi_j\Big]^{\frac{1}{2}}.\era\eeq We should also consider $\pm\lambda^2\sum\limits_{i,j=4}^{6}\nabla\phi_i.\nabla\phi_j\geq 0$ to get the lowest energy configuration. This should be found by canceling some of the terms in the second line of the expression given in (26). The simplest way is to require the orthogonality of $\nabla\phi_i$ and $\nabla\phi_j$. Then the lowest energy in the static gauge is \beq \tilde{\zeta}_3 = T_3 \int d^3\sigma \Big[ (1\pm \lambda^2
\stackrel{\rightarrow}{B}.\sum\limits_{i=4}^{6}\nabla \phi_i )^2
+\frac{1}{2} \sum\limits_{i,j=4}^{6} \mid \nabla \phi_i\wedge \nabla
\phi_j \mid^2 \Big]^{\frac{1}{2}}, \eeq with the constraints
\beq\sum\limits_{i=4}^{6}\nabla \phi_i =\pm
\stackrel{\rightarrow}{B},\eeq and the solution is similar to the
"bi-excited system", we find \beq\sum\limits_{i=4}^{6} \phi_i =\pm
\frac{N_m}{2r}.\eeq

The solution obtained for each excited scalar field has one collective
coordinate in the D3-brane world volume theory. This is the
direction along which the brane extends away from the D3-brane.
Thus, this collective coordinate represents a "ridge" solution in
the D3-brane theory.

Now, we will look at another case in which the electric field is
present, and to see how the energy of the system could be minimized
and what kind of solutions we could obtain.

\subsection{Addition of Electric Field}
\hspace{.3in}First we start by exciting two transverse directions
($\phi_4$ and $\phi_5$) with the electric field
$\stackrel{\rightarrow}{E}$ switched on. We consider as previous that
$\nabla\phi_4\bot\nabla\phi_5$. Then the energy of our system is
\beq\bra{lll} E_3&= T_3 \int d^3\sigma\Big[ 1 +\lambda^2 \Big( \mid
\nabla\phi_4 \mid^2 + \mid \nabla\phi_5 \mid^2 +\stackrel{\rightarrow}{B}^2 +\stackrel{\rightarrow}{E}^2
\Big)\\\\
&+\lambda^4 \Big[ (\stackrel{\rightarrow}{B}.\nabla\phi_4
)^2+(\stackrel{\rightarrow}{B}.\nabla\phi_5 )^2
+(\stackrel{\rightarrow}{E}.\stackrel{\rightarrow}{B})^2 +\mid
\nabla\phi_4 \wedge\nabla\phi_5\mid^2 +\mid
\stackrel{\rightarrow}{E}
\wedge\nabla\phi_4\mid^2 +\mid \stackrel{\rightarrow}{E} \wedge\nabla\phi_5\mid^2 \Big] \\\\
&+\lambda^6 \mid \stackrel{\rightarrow}{E}.(\nabla\phi_4
\wedge\nabla\phi_5 )\mid^2\Big]^{\frac{1}{2}}\\\\ &=T_3 \int
d^3\sigma\Big( \lambda^2 \mid \nabla\phi_4 + \nabla\phi_5
\pm(\stackrel{\rightarrow}{B}+\stackrel{\rightarrow}{E})\mid^2
+[1\mp\lambda^2 (\nabla\phi_4 +
\nabla\phi_5).(\stackrel{\rightarrow}{B}+\stackrel{\rightarrow}{E})]^2 \\\\
&- 2\lambda^2 B.E +\lambda^4(\stackrel{\rightarrow}{E}.\stackrel{\rightarrow}{B})^2 \\\\
&+ \lambda^4 \Big[ \mid \nabla\phi_4 \wedge\nabla\phi_5\mid^2 +\mid
\stackrel{\rightarrow}{E} \wedge\nabla\phi_4\mid^2 +\mid
\stackrel{\rightarrow}{E} \wedge\nabla\phi_5\mid^2
-(\stackrel{\rightarrow}{E}.\nabla\phi_4)^2
-(\stackrel{\rightarrow}{E}.\nabla\phi_5)^2\\\\
&-2\Big( \nabla\phi_4
.(\stackrel{\rightarrow}{E}+\stackrel{\rightarrow}{B})\nabla\phi_5
.(\stackrel{\rightarrow}{E}+\stackrel{\rightarrow}{B})+(\nabla\phi_4
.\stackrel{\rightarrow}{B})(\nabla\phi_4
.\stackrel{\rightarrow}{E})+ (\nabla\phi_5
.\stackrel{\rightarrow}{B})(\nabla\phi_5
.\stackrel{\rightarrow}{E})\Big) \Big]
\\\\ &+\lambda^6 \mid \stackrel{\rightarrow}{E}.(\nabla\phi_4 \wedge\nabla\phi_5
)\mid^2\Big)^{\frac{1}{2}}. \era\eeq By taking into consideration the
previous analysis in the absence of an electric field, the energy
$E_3$ becomes \beq\bra{lll} E_3&\geq T_3 \int d^3\sigma\Big(
[1\pm\lambda^2 (\nabla\phi_4 +
\nabla\phi_5).(\stackrel{\rightarrow}{B}+\stackrel{\rightarrow}{E})]^2
- 2\lambda^2 \stackrel{\rightarrow}{B}.\stackrel{\rightarrow}{E}
+\lambda^4(\stackrel{\rightarrow}{E}.\stackrel{\rightarrow}{B})^2\\\\
&+ \lambda^4 \Big[ \stackrel{\rightarrow}{E}^2
\mid\nabla\phi_4\mid^2 + \stackrel{\rightarrow}{E}^2
\mid\nabla\phi_5\mid^2 \Big] +\lambda^6 \mid
\stackrel{\rightarrow}{E}.(\nabla\phi_4 \wedge\nabla\phi_5
)\mid^2\Big)^{\frac{1}{2}}\era,\eeq This expression is consistent
with the fact that \beq\nabla\phi_4 + \nabla\phi_5
\pm(\stackrel{\rightarrow}{B}+\stackrel{\rightarrow}{E})=0\eeq and
\beq- 2\lambda^2 \stackrel{\rightarrow}{B}.\stackrel{\rightarrow}{E}
+\lambda^4(\stackrel{\rightarrow}{E}.\stackrel{\rightarrow}{B})^2\geq
0.\eeq We also used the following expression \beq\mid
\stackrel{\rightarrow}{E} \wedge\nabla\phi_5\mid^2
=\stackrel{\rightarrow}{E}^2 \mid \nabla\phi_5 \mid^2
-(\stackrel{\rightarrow}{E}.\nabla\phi_5)^2 .\eeq Then to get the
lowest energy in the presence of an electric field we require $-
2\lambda^2 \stackrel{\rightarrow}{B}.\stackrel{\rightarrow}{E}
+\lambda^4(\stackrel{\rightarrow}{E}.\stackrel{\rightarrow}{B})^2=
0$; i.e. $\stackrel{\rightarrow}{E}.\stackrel{\rightarrow}{B} =
\frac{2}{\lambda^2}$ or $\stackrel{\rightarrow}{E}\bot
\stackrel{\rightarrow}{B}$. With these simplifications, the energy
becomes \beq\bra{llll} \tilde{E}_3&= T_3 \int d^3\sigma\Big( \Big[
1\pm\lambda^2 (\nabla\phi_4 +
\nabla\phi_5).(\stackrel{\rightarrow}{B}+\stackrel{\rightarrow}{E})\Big]^2
+ \lambda^4 \stackrel{\rightarrow}{E}^2\Big[ \mid\nabla\phi_4\mid^2
+ \mid\nabla\phi_5\mid^2 \Big]\\\\&+\lambda^6 \mid
\stackrel{\rightarrow}{E}.(\nabla\phi_4 \wedge\nabla\phi_5
)\mid^2\Big)^{\frac{1}{2}}.\era\eeq 
In the following we consider $\stackrel{\rightarrow}{E}.\stackrel{\rightarrow}{B} =
\frac{2}{\lambda^2}$. We remark that the energy of D2$\bot$D3 brane is increased by the presence of the
electric field and by switching it off we obtain the lowest energy configuration obtained previously.

By solving (32) in the static gauge, using the Bianchi identity, we obtain the solution \beq\phi_4 +
\phi_5 = \mp\frac{ N_m +N_e}{2r},\eeq with $r^2 =\sum\limits_{a=1}^{3}(\sigma^a)^2$, $N_m$ and $N_e$ the magnetic and the electric charges respectively.

Now, by exciting three scalars, the energy is generalized to the
following \beq\bra{ll}
\xi_3&= T_3 \int d^3\sigma\Big[ 1+\lambda^2 ( \sum\limits_{i=4}^{6} \mid \nabla \phi_i \mid^2
+\stackrel{\rightarrow}{B}^2 +\stackrel{\rightarrow}{E}^2)\\\\
&+\lambda^4 \Big(
(\stackrel{\rightarrow}{E}.\stackrel{\rightarrow}{B})^2
+\sum\limits_{i=4}^{6} \mid \stackrel{\rightarrow}{B}.\nabla \phi_i
\mid^2 +\sum\limits_{i=4}^{6} \mid
\stackrel{\rightarrow}{E}\wedge\nabla \phi_i \mid^2 +
\frac{1}{2}\sum\limits_{i,j=4}^{6}
\mid \nabla \phi_i\wedge \nabla \phi_j \mid^2  \Big)\\\\
&+\lambda^6  \frac{1}{2}\mid
\stackrel{\rightarrow}{E}.(\sum\limits_{i,j=4}^{6}\nabla\phi_i
\wedge\nabla\phi_j )\mid^2 \Big]^{\frac{1}{2}} \era\eeq

We also consider as before $\nabla\Phi_i \bot\nabla\Phi_j$ and
$\stackrel{\rightarrow}{E}.\stackrel{\rightarrow}{B}=\frac{2}{\lambda^2}$
with $i,j=4,5,6$. Then the energy will be reduced to the lowest energy for
three excited directions \beq\bra{ll} \tilde{\xi}_3&= T_3 \int
d^3\sigma\Big( [1\pm\lambda^2 \sum\limits_{i=4}^{6}\nabla \phi_i
.(\stackrel{\rightarrow}{B}+\stackrel{\rightarrow}{E})]^2 +
\lambda^4 \sum\limits_{i,j=4}^{p}\stackrel{\rightarrow}{E}^2 ( \mid
\nabla \phi_i \mid^2
+ \mid \nabla\phi_j  \mid^2 )\\\\
&+\lambda^6 \mid
\stackrel{\rightarrow}{E}.(\sum\limits_{i,j=4}^{6}\nabla\phi_i
\wedge\nabla\phi_j )\mid^2\Big)^{\frac{1}{2}} ,\era\eeq with $i<j$ in
the summations and where we assume the following condition
\beq\sum\limits_{i=4}^{6}\nabla \phi_i \pm
(\stackrel{\rightarrow}{B}+\stackrel{\rightarrow}{E})=0.\eeq 
The last equation is easily solved to give the solution
\beq\sum\limits_{i=4}^{6} \phi_i =\mp \frac{ N_m + N_e}{2r}.\eeq
Again if we require that $\stackrel{\rightarrow}{E}$ is parallel to
one of $\nabla\phi_i$ the energy will be minimized
as the last term in the expression (38) of the energy vanishes. Then,
accordingly to (38) and (40) with
$$|\stackrel{\rightarrow}{B}|=\left|\frac{N_m}{2r^2}\right|,
\phantom{~~~~}|\stackrel{\rightarrow}{E}|=\left|\frac{N_e}{2r^2}\right|$$
the energy becomes \beq\bra{lll} \tilde{\xi}_3= T_3 \int
d^3\sigma\Big( [1+\lambda^2 \frac{(N_m +N_e)^2}{4r^4}]^2 + \lambda^4
\frac{N_e^2 (N_m +N_e)^2}{8r^6} \Big)^{\frac{1}{2}} \era\eeq This is
the lowest energy for both cases where two or three transverse
scalars are excited since the solution is a superposition
of the scalars.

\section{Dyonic and Non-Static D1$\bot$D3 Funnels}
\hspace{.3in}In the D1 worldvolume point of view, the intersection of D1 and D3-brane is described as a funnel of increasing radius as we approach the D3 brane, where the D-strings expand into a fuzzy two-sphere. In the D3-picture the worldvolume solution includes a BPS magnetic monopole and the Higgs field is interpreted as a transverse spike. To enlarge the discussion of this subject, we lift the static condition and we keep the electric field on.

\subsection{Space-Time dependent Solution}
We start by D1-picture. The action is given by the non-abelian BI action of ($N-N_f$) strings. By considering again the ansatz (9) the action (6) became
\beq
S=-T_1\int d^2\sigma STr \sqrt{1+\lambda^2 c\hat{R}^{'2}-\lambda^2 c\dot{\hat{R}}^2 -\lambda^2 E^2}\sqrt{1+4\lambda^2 c \hat{R}^4},
\eeq
with the notations $\hat{R}'=\partial_{\sigma}\hat{R}$ and $\dot{\hat{R}}=\partial_{\tau}\hat{R}$. By varying this with respect to $\hat{R}$ we recover the full equations of motion given by
\beq
2\lambda^2 c\dot{\hat{R}}\hat{R}'\dot{\hat{R}}' \hat{R}'' (1-\lambda^2 c\dot{\hat{R}}^2)-\ddot{\hat{R}}(1+\lambda^2 c\hat{R}^{'2})=8\hat{R}^3 (\frac{1+\lambda^2 c\hat{R}^{'2}-\lambda^2 c\dot{\hat{R}}^2 -\lambda^2 E^2}{1+4\lambda^2 c \hat{R}^4}).
\eeq 
To get these equations in terms of dimensionless variables we consider the following rescalings
$$
r=\sqrt{2\lambda\sqrt{c}}\hat{R}\phantom{~~~~}\tilde\tau=\sqrt{\frac{2}{\lambda\sqrt{c}}}\tau\phantom{~~~~}\tilde\sigma=\sqrt{\frac{2}{\lambda\sqrt{c}}}\sigma.
$$
Then the above equations of motion can be written in a Lorentz-invariant form
\beq
\partial_\mu \partial^\mu r +(\partial_\mu \partial^\mu r)(\partial_\nu r)(\partial^\nu r)-(\partial_\mu \partial^\nu r)(\partial_\nu r)(\partial^\mu r)=2r^3 \frac{1+ (\partial_\mu r)(\partial^\mu r)-\lambda^2 E^2}{1+r^4},
\eeq 
where $\mu$ and $\nu$ can take the values $\tilde\tau$ and $\tilde\sigma$. The re-scaled action and energy density of the configuration are
\beq
\tilde{S}=-T_1\int d^2\sigma STr \sqrt{1+r'^2-\dot{r}^2 -\lambda^2 E^2}\sqrt{1+r^4},
\eeq 
\beq
T_{\tau\tau}=E=(1+r'^2 -\lambda^2 E^2)\sqrt{\frac{1+r^4}{1+r'^2-\dot{r}^2 -\lambda^2 E^2}},
\eeq
and the pressure is given by
\beq
T_{\sigma\sigma}=(1-\dot{r}^2 -\lambda^2 E^2)\sqrt{\frac{1+r^4}{1+r'^2-\dot{r}^2 -\lambda^2 E^2}},
\eeq
with dots and primes implying differentiation with respect to the re-scaled time and space respectively. 

Now if we consider $r_0$ is the initial radius of the collapsing configuration where $\dot{r}=0$, the energy is $E=(1+r'^2_0 -\lambda^2 E^2) \sqrt{\frac{1+r^4_0}{1+r'^2_0-\lambda^2 E^2}}$. The conserved energy density leads then at large $N$ to the following equation
\beq
\dot{r}^2=(1+r'^2-\lambda^2 E^2)-\frac{1+r^4}{1+r^4_0 }\Big( \frac{1+r'^2 -\lambda^2 E^2}{1+r'^2_0 -\lambda^2 E^2}\Big)^2 (1+r'^2_0-\lambda^2 E^2 ).
\eeq

In purely time dependence, we get
\beq
\dot{r}^2=(1-\lambda^2 E^2)\frac{r^4_0-r^4}{1+r^4_0}.
\eeq
Then we can write
\beq
\int\limits_{0}^{t}dt= \sqrt{\frac{1+r^4_0}{1-\lambda^2 E^2}}\int\limits_{r_0}^{r}\frac{dr}{\sqrt{r^4_0-r^4}}.
\eeq
Now if we consider the initial radius at $t=0$ is $r_1$ at $\dot{r}=u$, the equation (48) becomes
\beq
\dot{r}^2=\frac{r^4_1 (1-\lambda^2 E^2)-r^4 (1-u^2-\lambda^2 E^2)+u^2}{1+r^4_1},
\eeq
and we can solve the following equality
\beq
\int\limits_{0}^{t}dt= \sqrt{\frac{1+r^4_1}{1-u^2-\lambda^2 E^2}}\int\limits_{r_1}^{r}\frac{dr}{\sqrt{r^4_1-r^4}}.
\eeq
we get
\beq
it\sqrt{\frac{1-u^2-\lambda^2 E^2}{1+r^4_1}}= \int\limits_{r_1}^{r_2}\frac{dr}{\sqrt{r^4-r^4_1}}+\int\limits_{r_2}^{r}\frac{dr}{\sqrt{r^4-r^4_1}}.
\eeq
Then this leads to
\beq
i\tilde{t}=it\sqrt{2}\sqrt{\frac{1-u^2-\lambda^2 E^2}{1+r^4_1}}r_2=cn^{-1}(\frac{r_1}{r_2},\frac{1}{\sqrt{2}})+cn^{-1}(\frac{r_2}{r_1},\frac{1}{\sqrt{2}}).
\eeq
Let $T=cn^{-1}(\frac{r_1}{r_2},\frac{1}{\sqrt{2}})$ the the solution is
\beq
r(t)=r_2 cn(\tilde{t}+T,\frac{1}{\sqrt{2}}).
\eeq
If $u=0$ we find
\beq
it\sqrt{2}\sqrt{\frac{1-\lambda^2 E^2}{1+r^4_1}}r_1=cn^{-1}(\frac{r_1}{r},\frac{1}{\sqrt{2}}).
\eeq
Then if $E=0$ we get
\beq
\dot{r}^2=\frac{r^4_1 -r^4 (1-u^2)}{1+r^4_1},
\eeq
and
\beq
it\sqrt{2}\sqrt{\frac{1-\lambda^2 E^2}{1+r^4_1}}r_2=cn^{-1}(\frac{r_1}{r_2},\frac{1}{\sqrt{2}})+cn^{-1}(\frac{r_2}{r_1},\frac{1}{\sqrt{2}}).
\eeq
The last case is $E=0$ and $u=0$ and this was discussed in the reference \cite{brane}. For the first and the second cases, we remark that if $r_1\rightarrow\infty$ we get $\dot{r}^2=1-\lambda^2 E^2$ but for the third and the forth cases we have $\dot{r}^2=1$, so for at the presence of electric field the brane collapses lesser that the speed of light. For the both cases $E=0$ and $E\ne0$ a collapsing speed does not depend on the initial speed.

In the purely spatial independence we follow the same way as time dependence. Let's consider at $t=0$, $r'=v$ and $\dot{r}=0$ and the initial radius is $r_1$ and the purely spatial solution for $T_{\sigma\sigma}$ is gotten by solving the following, first we have
\beq
r'^2=\frac{r^4 (1+v^2-\lambda^2 E^2) -r^4_1(-\lambda^2 E^2)+v^2}{1+r^4_1},
\eeq
and
\beq
\int\limits_{0}^{\sigma}d\sigma= \sqrt{\frac{1+r^4_1}{1-v^2-\lambda^2 E^2}}\int\limits_{r_1}^{r}\frac{dr}{\sqrt{r^4-r'^4_1}},
\eeq
where $r'^4_1 =r_1^4 -\frac{v^2 (1+r_1^4 )}{1+v^2 -\lambda^2 E^2}$ and $\int\limits_{r_1}^{r}=\int\limits_{r'_1}^{r}-\int\limits_{r'_1}^{r_1}$ so we find
\beq
\frac{\sigma\sqrt{1+v^2 -\lambda^2 E^2}}{\sqrt{1+r_1^4 }}= \int\limits_{r'_1}^{r}\frac{dr}{\sqrt{r^4-r'^4_1}}-\int\limits_{r'_1}^{r_1}\frac{dr}{\sqrt{r^4-r'^4_1}},
\eeq
then we get
\beq\bra{ll}
\frac{\sigma\sqrt{2}r'_1 \sqrt{1+v^2 -\lambda^2 E^2}}{\sqrt{1+r_1^4 }}&=cn^{-1}(\frac{r'_1}{r},\frac{1}{\sqrt{2}})-cn^{-1}(\frac{r'_1}{r_1},\frac{1}{\sqrt{2}}),\\\\
&=cn^{-1}\Big( \frac{\frac{r'^2_1}{rr_1}+\frac{1}{2}\sqrt{(1-(\frac{r'_1}{r})^4)(1-(\frac{r'_1}{r_1})^4)}}{1+(1-r'^2_1)(1-r_1^2)},\frac{1}{\sqrt{2}}\Big)
\era
\eeq
For $v^2 =r_1^4 (1-\lambda^2 E^2)$ we get $$r'^2 =r^4 (1-\lambda^2 E^2)$$ which leads to the following
\beq
\mp r=\frac{1}{\sigma(1-\lambda^2 E^2)\mp \frac{1}{r}}.
\eeq
Thus for particular initial conditions of $r'$ and $r$, we do not find the infinite periodic
brane-anti-brane array, but rather a solution with a decaying r. In the presence of an
electric field it is attenuated by the factor of $1/\sqrt{1-\lambda^2E^2}$. The equation $r_0^2=r^4(1-\lambda^2E^2)$ provides a generalization of the BPS equation in presence of an electric field. Note that for $E = 0$ we get the usual BPS equation $r_0^2 = r^4$. The BPS equation can be
obtained by studying the condition to have an unbroken supersymmetry, which amounts
to the condition for the vanishing of the variation of gaugino on the world volume of the
intersection \cite{timeDepen}.

For equations (51) and (59), consider the Wick rotation $\tau\rightarrow i\sigma$. This provides
interesting results in the cases where $E$, $u$ and $v$ are all zero. Here we find $(51)\rightarrow (59)$ for $\tau\rightarrow i\sigma$ and $u\rightarrow iv$. Thus in the presence of the electric field, the space time dependent solutions generalize nicely.
\subsection{Automorphisms}
Another interesting feature of these solutions is the $r\leftrightarrow1/r$ duality, which arises as the
consequence of the invariance of the complex curve under an $r\leftrightarrow 1/r$ automorphism. More precisely, in the absence of any worldvolume electric field and with $u = 0$, we find (51) reduces to the equation of a complex curve $s^2 =\frac{r_0^4 -r^4}{1+r^4}$  where we have defined $\dot{r}= s$ and $r$ and $s$ are interpreted as complex variables. For this curve the relevant automorphisms have been studied in \cite{timeDepen} and a connection to the $\tau\rightarrow i\sigma$ duality has been made. Here we find a similar duality. Towards this end, we study an automorphism of the curve (51) for $E\ne 0$ and $v\ne 0$. After defining $s = \dot{r}$ we find
\beq
s^2=\frac{r^4_1 (1-\lambda^2 E^2)-r^4 (1-u^2-\lambda^2 E^2)+u^2}{1+r^4_1}.
\eeq

The automorphism acts as $r\rightarrow R = 1/r$ , $r_0 \rightarrow R_0 = 1/r_0$, $s \rightarrow =\tilde{s} =is/r^2$, $u\rightarrow u' = -iuR^2$. This automrphism acts at fixed $E$. For $u = 0$, $r\rightarrow R = r^2_0/r$, $s \rightarrow =\tilde{s} =isr^2_0/r^2$, is still a good automorphism of (64).

\section{Dyonic Fluctuations}
\hspace{.3in}In this section, we give an examination of the propagation of the fluctuations on the fuzzy funnel. The setup is similar to both D1$\bot$D5 and D1$\bot$D3 systems. We notice that there are two basic types of funnel's fluctuations, the overall transverse ones in the directions perpendicular to both the Dp-brane (p=3,5) and the string (i.e., $X^{p+1,..,8}$), and the relative transverse ones which are transverse to the string, but parallel to the Dp-brane world volume (i.e., along X$^{1,..,p}$).

Thus, we treat the dynamics of the funnel solutions. We solve the linearized equations of motion for small and
time-dependent fluctuations of the transverse scalars around the exact background in dyonic case.
\subsection{Overall Transverse Fluctuations in D1$\bot$D3 System}
\subsubsection{Zero Mode}
We deal with the fluctuations of the funnel (10) discussed in the previous section. By plugging into the full ($N,N_f$)-string action
(6,7) the "overall transverse" $\delta \phi^m (\sigma,t)=f^m (\sigma,t)I_N$, $m=4,...,8$ which is the simplest type of fluctuation with $I_N$ the identity matrix, together with the funnel solution, we get \beq\bra{llll} S&=-T_1\int d^2\sigma STr \Big[
(1+\lambda E)(1+\frac{\lambda^2 \al^i \al^i}{4(1-\lambda^2E^2)^2\sigma^4 }) \\&\phantom{~~~~~~~~~}\Big( (1+\frac{\lambda^2 \al^i \al^i}{4(1-\lambda^2E^2)^2\sigma^4})(1-(1-\lambda E)\lambda^2 (\partial_t \delta\phi^m )^2) +\lambda^2 (\partial_\sigma \delta\phi^m )^2\Big) \Big]^{1\over 2}\\\\
&\approx-NT_1\int d^2\sigma H \Big[ (1+\lambda E)-(1-\lambda^2 E^2)\frac{\lambda^2}{2} (\dot{f}^m)^2 +\frac{(1+\lambda E)\lambda^2}{2H} (\partial_\sigma f^m)^2 +...\Big] \era \eeq where $$H=1+\frac{\lambda^2 C}{4(1-\lambda^2E^2)^2\sigma^4}$$ and $C=Tr \al^i \al^i$. For
the irreducible $N\times N$ representation we have $C=N^2 -1$. In the last line we have only kept the terms quadratic in the fluctuations as this is sufficient to determine the linearized equations of motion \beq\Big((1-\lambda E)(1+\lambda^2\frac{N^2-1}{4(1-\lambda^2E^2)^2\sigma^4})\partial^{2}_{t}-\partial^{2}_{\sigma}\Big) f^m =0.\eeq 

In the overall case, all the points of the fuzzy funnel move or fluctuate in the same direction of the dyonic string by an equal distance $\delta x^m$. Thus, the fluctuations $f^m$ could be rewritten as follows \beq f^m (\sigma,t)=\Phi(\sigma)e^{-iwt}\delta x^m,\eeq where $\Phi$ is a function of the spatial coordinate. With this ansatz the equation of motion (66) becomes \beq\Big( (1-\lambda E)(1+\lambda^2\frac{N^2 -1}{4(1-\lambda^2E^2)^2\sigma^4}) w^2 +\partial^{2}_{\sigma}\Big) \Phi(\sigma)=0.\eeq Then, the problem is reduced to finding the solution of a single scalar equation.

We consider the physical phenomenon which is defined by the fact that the electric field $E$ is in the interval $[0, \frac{1}{\lambda}[$ (contrary to what was treated in \cite{fluc}, such that $E$ was tending to $\infty$).

The equation (68) is an analog one-dimensional Schr\"odinger equation. Let's rewrite it as \beq\Big(\frac{1}{w^2(1-\lambda E)}\partial^{2}_{\sigma}+1+\frac{\lambda^2 N^2 }{4(1-\lambda^2E^2)^2\sigma^4}\Big) \Phi(\sigma)=0,\eeq for large $N$. If we suggest \beq\tilde{\sigma}=w\sqrt{1-\lambda E}\sigma,\eeq the equation (69) becomes \beq\Big(\partial^{2}_{\tilde{\sigma}}+1+\frac{\kappa^2}{\tilde{\sigma}^4}\Big) \Phi(\tilde{\sigma})=0,\eeq with the potential is \beq V(\tilde{\sigma})=\frac{\kappa^2}{\tilde{\sigma}^4}, \eeq and \beq \kappa=\frac{\lambda N w^2}{2(1+\lambda E)}.\eeq The equation (71) is a Schr\"odinger equation for an attractive singular potential $\propto\tilde{\sigma}^{-4}$ and depends on the single coupling parameter $\kappa$ with constant positive Schr\"odinger energy. The solution is then known by making the following coordinate change \beq \chi(\tilde{\sigma})=\int\limits^{\tilde{\sigma}}_{\sqrt{\kappa}} dy\sqrt{1+\frac{\kappa^2}{y^4}}, \eeq and \beq \Phi=(1+\frac{\kappa^2}{\tilde{\sigma}^4})^{-\frac{1}{4}}\tilde{\Phi}. \eeq Thus, the equation (71) becomes \beq\Big( -\partial^{2}_{\chi}+V(\chi)\Big) \tilde{\Phi}=0,\eeq with \beq V(\chi)=\frac{5\kappa^2}{(\tilde{\sigma}^2+\frac{\kappa^2}{\tilde{\sigma}^2})^3}.\eeq

Accordingly to the variation of this potential (Fig.1), the system looks like separated into two regions depending on $\sigma$. In small $\sigma$ region $V$ is close to 0 with a constant value for all $E$. In large $\sigma$ region, specially when $\sigma$ reaches 0.7, $V$ increases too fast as we jump to a new region and gets a maximum value when $E\approx0.5$. 

Then, the fluctuation is found to be \beq \Phi=(1+\frac{\kappa^2}{\tilde{\sigma}^4})^{-\frac{1}{4}}e^{\pm i\chi(\tilde{\sigma})}. \eeq This fluctuation has the following limits; at large $\sigma$, $\Phi\sim e^{\pm i\chi(\tilde{\sigma})}$ and if $\sigma$ is small
$\Phi=\frac{\tilde{\sigma}}{\sqrt{\kappa}}e^{\pm i\chi(\tilde{\sigma})}$. These are the asymptotic wave function in the regions $\chi\rightarrow \pm\infty$, while around $\chi\sim 0$; i.e. $\tilde{\sigma}\sim\sqrt{\kappa}$, $\Phi\sim 2^{-\frac{1}{4}}$. Also we find that $\Phi$ has different expressions in small and large $\sigma$ regions.
\subsubsection{Non-Zero Modes}
The fluctuations discussed above could be called the zero mode $\ell=0$ and for non-zero modes $\ell\geq0$, the fluctuations are $\delta \phi^m (\sigma,t)=\sum\limits^{N-1}_{\ell=0}\psi^{m}_{i_1 ... i_\ell}\al^{i_1} ... \al^{i_\ell} $ with $\psi^{m}_{i_1 ... i_\ell}$ are completely symmetric and traceless in the lower indices.

The action describing this system is \beq\bra{lll}
S&\approx-NT_1\int d^2\sigma  \Big[ (1+\lambda E)H-(1-\lambda^2
E^2)H\frac{\lambda^2}{2} (\partial_{t}\delta\phi^m)^2) \\\\
&+\frac{(1+\lambda E)\lambda^2}{2H} (\partial_\sigma \delta\phi^m)^2
-(1-\lambda^2 E^2)\frac{\lambda^2}{2}\lbrack \phi^i ,\delta\phi^m
\rbrack^2 \\\\
&-\frac{\lambda^4}{12}\lbrack \partial_{\sigma}\phi^i
,\partial_{t}\delta\phi^m \rbrack^2+...\Big] \era\eeq Now the
linearized equations of motion are \beq \Big[(1+\lambda
E)H\partial_{t}^2 -\partial_{\sigma}^2\Big]\delta\phi^m
+(1-\lambda^2 E^2)\lbrack \phi^i ,\lbrack \phi^i ,\delta\phi^m
\rbrack\rbrack -\frac{\lambda^2}{6}\lbrack \partial_{\sigma}\phi^i
,\lbrack \partial_{\sigma}\phi^i ,\partial^2 _{t}\delta\phi^m
\rbrack\rbrack=0,\eeq with $H=1+\lambda^2\frac{N^2 -1}{4(1-\lambda^2E^2)^2\sigma^4}$. Since the background solution is $\phi^i
\propto \al^i$ and we have $\lbrack \al^i , \al^j \rbrack
=2i\epsilon_{ijk}\al^k $, we get \beq\bra{ll} \lbrack \al^i ,
\lbrack\al^i, \delta\phi^m \rbrack
&=\sum\limits_{\ell<N}\psi^{m}_{i_1 ... i_\ell}\lbrack \al^i ,
\lbrack\al^i ,\al^{i_1} ... \al^{i_\ell} \rbrack\\\\
&=\sum\limits_{\ell<N}4\ell(\ell+1)\psi^{m}_{i_1 ...
i_\ell}\al^{i_1} ... \al^{i_\ell} \era\eeq To obtain a specific
spherical harmonic on 2-sphere, we have \beq\lbrack \phi^i ,\lbrack
\phi^i ,\delta\phi_{\ell}^m
\rbrack\rbrack=\frac{\ell(\ell+1)}{(1-\lambda^2E^2)\sigma^2}\delta\phi_{\ell}^m
,\phantom{~~~~~~}\lbrack \partial_{\sigma}\phi^i ,\lbrack
\partial_{\sigma}\phi^i ,\partial_{t}^2 \delta\phi^m
\rbrack\rbrack=\frac{\ell(\ell+1)}{(1-\lambda^2E^2)^2\sigma^4}\partial_{t}^2\delta\phi
_{\ell}^m .\eeq Then for each mode the equations of motion are \beq
\Big[ \Big( (1+\lambda E)(1+\lambda^2\frac{N^2 -1}{4(1-\lambda^2E^2)^2\sigma^4})
-\frac{\lambda^2\ell(\ell+1)}{6(1-\lambda^2E^2)^2\sigma^4}\Big) \partial_{t}^2
-\partial_{\sigma}^2 +\frac{\ell(\ell+1)}{\sigma^2}
\Big]\delta\phi_{\ell}^m =0.\eeq The solution of the equation of
motion can be found by taking the following proposal. Let's consider
$\phi_{\ell}^m =f^m_\ell (\sigma)e^{-iwt}\delta x^m$ in direction
$m$ with $f^m_\ell (\sigma)$ is some function of $\sigma$ for each
mode $\ell$.

The last equation can be rewritten as \beq \Big[-\partial_{\sigma}^2
+V(\sigma) \Big] f_{\ell}^m (\sigma)=w^2 (1+\lambda E) f_{\ell}^m
(\sigma),\eeq with
$$V(\sigma)=-w^2 \Big( (1+\lambda E)\frac{\lambda^2 N^2}{4(1-\lambda^2E^2)^2\sigma^4}
-\frac{\lambda^2\ell(\ell+1)}{6(1-\lambda^2E^2)^2\sigma^4}\Big)+\frac{\ell(\ell+1)}{\sigma^2}.$$

In small $\sigma$ region, this potential is reduced to
$$V(\sigma)= \frac{-w^2\lambda^2}{(1-\lambda^2E^2)^2\sigma^4}\Big( \frac{ (1+\lambda E)N^2}{4}
-\frac{\ell(\ell+1)}{6}\Big).$$
This potential (Fig.2(a)) is close to 0 for almost of $\sigma$ and $E$ until that $E\approx0.87$ we remark that $V$ changes at $\sigma\approx0.04$ and then goes up too fast to be close to 0 again for the other values of $\sigma$.

In small $\sigma$ limit, we reduce the equation (84) to the following form \beq \Big[ w^2 \Big( (1+\lambda E)(1+\lambda^2\frac{N^2 -1}{4(1-\lambda^2E^2)^2\sigma^4})
-\frac{\lambda^2\ell(\ell+1)}{6(1-\lambda^2E^2)^2\sigma^4}\Big)+\partial_{\sigma}^2 \Big]
f_{\ell}^m  (\sigma)= 0.\eeq and again as \beq \Big[ 1+
\frac{1}{(1-\lambda^2E^2)^2\sigma^4}\Big(\lambda^2\frac{N^2
-1}{4}-\frac{\lambda^2\ell(\ell+1)}{6(1+\lambda E)}\Big)+\frac{1}{w^2 (1+\lambda
E)}\partial_{\sigma}^2  \Big] f_{\ell}^m  (\sigma)= 0.\eeq
We define new coordinate $\tilde{\sigma}=w\sqrt{1+\lambda E}\sigma$
and the latter equation becomes \beq
\Big[\partial_{\tilde{\sigma}}^2 + 1+
\frac{\kappa^2}{\tilde{\sigma}^4}\Big] f_{\ell}^m (\sigma)= 0,\eeq where
$$
\kappa^2=\frac{w^2(1+\lambda E)}{(1-\lambda^2E^2)^2}\Big(\lambda^2\frac{N^2
-1}{4}-\frac{\lambda^2\ell(\ell+1)}{6(1+\lambda
E)}\Big)^{\frac{1}{2}}$$
such that $$N>\sqrt{\frac{2\ell(\ell+1)}{3(1+\lambda E)}+1}.$$
By following the same setup of zero mode, we get the solution by using the steps
(74-78) with new $\kappa$. Since we considered small $\sigma$ we get
\beq
V(\chi)=\frac{5\tilde{\sigma}^6}{\kappa^4},
\eeq
and the fluctuation is found to be \beq f^m_\ell=\frac{\tilde{\sigma}}{\sqrt{\kappa}}e^{\pm
i\chi(\tilde{\sigma})}.\eeq in small $\sigma$ region.

Now, let's check the case of large $\sigma$. In this case, the equation of motion (84) of the fluctuation can be rewritten in the following form \beq \Big[-\partial_{\sigma}^2 +V(\sigma) \Big] f_{\ell}^m (\sigma)= w^2 (1+\lambda E) f_{\ell}^m (\sigma),\eeq with
$$V(\sigma)=\frac{\ell(\ell+1)}{\sigma^2}.$$ We remark that, in large $\sigma$ limit (Fig.2(b)), the potential $V$ is independent of $E$ and going down as $\sigma$ is going up. The figures 2(a) and 2(b) show that the system in non-zero modes is separated to two totally different regions and the main remark is that the potential gets a singularity at some level of $\sigma$ which is considered the intersection of small and large $\sigma$ regions. In our calculations we took small $\sigma$ from zero until the half of the unit of $\lambda=1$ and the large $\sigma$ region from 0.5 until 1 with $w=1$, $l=1$ and $N=10$.

The $f_{\ell}^m$ is now a Sturm-Liouville eigenvalue problem. The fluctuation is found to be \beq\bra{ll}f^m_\ell(\sigma)&=\alpha \sqrt{\sigma}BesselJ\Big(\frac{1}{2}\sqrt{1+4\ell(\ell+1)},w\sigma\sqrt{1+\lambda E}\Big)\\ &+\beta\sqrt{\sigma}BesselY\Big(\frac{1}{2}\sqrt{1+4\ell(\ell+1)},w\sigma\sqrt{1+\lambda E}\Big),\era\eeq with $\alpha$, $\beta$ are constants. Again, it's clear that the fluctuation solution in this case is totally different from the one gotten in small $\sigma$ limit (89) supporting the idea that the system is divided to two regions. In the following, we continue the study of D1$\bot$D3 branes by dealing with the relative transverse fluctuations.

\subsection{Relative Transverse Fluctuations in D1$\bot$D3 System}
\subsubsection{Zero Mode}
\hspace{.3in}In this subsection, we consider the "relative transverse" fluctuations $\delta
\phi^i (\sigma,t)=f^i (\sigma,t)I_N$, $i=1,2,3$, and the action describing the system has the expression \beq S=-T_1\int d^2\sigma STr \Big[ -det\pmatrix{\eta_{ab}+\lambda F_{ab}& \lambda \partial_a (\phi^j +\delta\phi^j )\cr -\lambda \partial_b (\phi^i +\delta\phi^i )& Q^{ij}_*
\cr}\Big]^{1\over2},\eeq with $$Q^{ij}_* =Q^{ij}+i\lambda (\lbrack \phi_i,\delta\phi_j \rbrack+\lbrack \delta\phi_i,\phi_j
\rbrack+\lbrack \delta\phi_i,\delta\phi_j \rbrack).$$ As done above, we keep only the terms quadratic in the fluctuations and the action becomes \beq S\approx-NT_1\int d^2\sigma  \Big[ (1-\lambda^2
E^2)H-(1-\lambda E)\frac{\lambda^2}{2}(\dot{f}^i)^2+\frac{(1+\lambda E)\lambda^2}{2H} (\partial_\sigma
f^i)^2 +...\Big],\eeq with $H=(1+\lambda^2\frac{N^2 -1}{4(1-\lambda^2E^2)^2\sigma^4})$.

Then we define the relative transverse fluctuation as $f^i =\Phi^i (\sigma)e^{-iwt}\de x^i$ in the direction of $x^i$, with $\Phi$ is a function of $\sigma$, and the equations of motion of the fluctuations are found to be \beq\Big ( -\partial^{2}_{\sigma}-\frac{w^{2}\lambda^2(1-\lambda E)(N^2-1)}{4(1+\lambda E)(1-\lambda^2E^2)^2\sigma^4}\Big)\Phi^i = w^2 \frac{1-\lambda E}{1+\lambda E}\Phi^i,\eeq where the potential is $$V(\sigma)=-\frac{w^{2}\lambda^2(1-\lambda E)(N^2-1)}{4(1+\lambda E)(1-\lambda^2E^2)^2\sigma^4}.$$ We remark that the presence of E is quickly increasing the potential from $-\infty$ to zero. Then, when $E$ is close to the inverse of $\lambda$ the potential is close to zero for all $\sigma$;
\begin{itemize} \item $E\sim 0$, $V(\sigma)\sim -\lambda^2\frac{N^2
-1}{4\sigma^4}w^{2}$ \item $E\sim \frac{1}{\lambda}$, $V(\sigma)\sim
-\frac{1-\lambda E}{2}\lambda^2\frac{N^2-1}{4(1-\lambda^2E^2)^2\sigma^4}w^{2}$.
\end{itemize}
This case is seen as a zero mode of what is following so we will focus on its general case known as non-zero modes.
\subsubsection{Non-Zero Modes}
\hspace{.3in}Let's give the equation of motion of relative transverse fluctuations of non-zero $\ell$ modes with ($N,N_f$)-strings intersecting D3-branes. The fluctuation is given by $\delta \phi^i
(\sigma,t)=\sum\limits^{N-1}_{\ell=1}\psi^{i}_{i_1 ...i_\ell}\al^{i_1} ... \al^{i_\ell} $ with $\psi^{i}_{i_1 ... i_\ell}$ are completely symmetric and traceless in the lower indices.

The action describing this system is \beq\bra{lll}
S&\approx-NT_1\int d^2\sigma  \Big[ (1-\lambda^2 E^2 )H-(1-\lambda
E)H\frac{\lambda^2}{2} (\partial_{t}\delta\phi^i)^2) \\\\
&+\frac{(1+\lambda E)\lambda^2}{2H} (\partial_\sigma \delta\phi^i)^2
-(1-\lambda E)\frac{\lambda^2}{2}\lbrack \phi^i ,\delta\phi^i
\rbrack^2 \\\\
&-\frac{\lambda^4}{12}\lbrack \partial_{\sigma}\phi^i
,\partial_{t}\delta\phi^i \rbrack^2+...\Big]. \era\eeq The equation of motion for relative transverse fluctuations in non-zero modes is
\beq \Big[\frac{1-\lambda E}{1+\lambda E}H\partial_{t}^2
-\partial_{\sigma}^2\Big]\delta\phi^i +(1-\lambda E)\lbrack \phi^i
,\lbrack \phi^i ,\delta\phi^i \rbrack\rbrack
-\frac{\lambda^2}{6}\lbrack \partial_{\sigma}\phi^i ,\lbrack
\partial_{\sigma}\phi^i ,\partial^2 _{t}\delta\phi^i
\rbrack\rbrack=0.\eeq
By the same way followed in overall case the equation of motion for each mode $\ell$ is found to be \beq \Big[
-\partial_{\sigma}^2+\Big( \frac{1-\lambda E}{1+\lambda
E}(1+\lambda^2\frac{N^2 -1}{4(1-\lambda^2E^2)^2\sigma^4})-\frac{\lambda^2
\ell(\ell+1)}{6(1-\lambda^2E^2)^2\sigma^4}  \Big ) \partial_{t}^2 +\frac{\ell(\ell+1)}{(1+\lambda
E)\sigma^2}\Big]\delta\phi^i_\ell =0. \eeq We
write $\de\phi^i_\ell =f^i_\ell e^{-iwt}\de x^i$ in the direction of $x^i$, then the equation (97) becomes \beq \Big[ -\partial_{\sigma}^2 -\Big( \frac{1-\lambda
E}{1+\lambda E}(1+\lambda^2\frac{N^2 -1}{4(1-\lambda^2E^2)^2\sigma^4})-\frac{\lambda^2
\ell(\ell+1)}{6(1-\lambda^2E^2)^2\sigma^4}  \Big ) w^2 +\frac{\ell(\ell+1)}{(1+\lambda
E)\sigma^2}\Big]  f^i_\ell =0. \eeq To solve this equation we start, for simplicity, by considering small $\sigma$. The equation (98) is reduced to \beq \Big[
-\partial_{\sigma}^2 -\frac{\lambda^2w^2}{(1-\lambda^2E^2)^2\sigma^4}\Big(\frac{1-\lambda E}{1+\lambda E}\frac{N^2 -1}{4}-\frac{\ell(\ell+1)}{6} \Big )  \Big] f^i_\ell = \frac{1-\lambda E}{1+\lambda E}w^2f^i_\ell , \eeq 
with the potential $$V= \frac{-\lambda^2w^2}{(1-\lambda^2E^2)^2\sigma^4}\Big(\frac{1-\lambda E}{1+\lambda E}\frac{N^2 -1}{4}-\frac{\ell(\ell+1)}{6} \Big ).$$ The potential $V$ is quite zero for all $E$ and only at $\sigma\approx0.02$ that we see $V$ varies in terms of $E$ and goes up too fast to be close to zero as a constant function (Fig.4(a)).

The equation of motion (99) can be rewritten as follows \beq \Big[ -\frac{1+\lambda E}{1-\lambda
E}\partial_{\sigma}^2 -\Big( (1+\lambda^2\frac{N^2-1}{4(1-\lambda^2E^2)^2\sigma^4})-\frac{1+\lambda E}{1-\lambda E}\frac{\lambda^2\ell(\ell+1)}{6(1-\lambda^2E^2)^2\sigma^4}  \Big ) w^2 \Big] f^i_\ell = 0. \eeq We change the coordinate to $\tilde{\sigma}=\sqrt{\frac{1-\lambda E}{1+\lambda E}}w\sigma$ and the equation (100) is rewritten as \beq \Big[\partial_{\tilde{\sigma}}^2+ 1+ \frac{\kappa^2}{\tilde{\sigma}^4}\Big] f^i_\ell (\tilde{\sigma})=0,\eeq with $$\kappa^2=w^4\lambda^2 \frac{3(1-\lambda E)^2 (N^2-1)-2(1-\lambda^2E^2)\ell(\ell+1)}{12(1+\lambda E)^2(1-\lambda^2E^2)^2}.$$ Then we follow the suggestions of WKB by making a coordinate change; \beq
\beta(\tilde{\sigma})=\int\limits^{\tilde{\sigma}}_{\sqrt{\kappa}}dy\sqrt{1+
\frac{\kappa^2}{y^4}}, \eeq and \beq f^i_\ell
(\tilde{\sigma})=(1+\frac{\kappa^2}{\tilde{\sigma}^4})^{-\frac{1}{4}}\tilde{f}^i_\ell
(\tilde{\sigma}). \eeq Thus, the equation (101) becomes \beq\Big(
-\partial^{2}_{\beta}+V(\beta)\Big) \tilde{f^i}=0,\eeq with \beq
V(\beta)=\frac{5\kappa^2}{(\tilde{\sigma}^2
+\frac{\kappa^2}{\tilde{\sigma}^2})^3}. \eeq Then \beq
f^i_\ell=(1+\frac{\kappa^2}{\tilde{\sigma}^4})^{-\frac{1}{4}}e^{\pm
i\beta(\tilde{\sigma})}. \eeq Since we are dealing with small $\sigma$ case the obtained fluctuation becomes $$f^i_\ell=\frac{\tilde{\sigma}}{\sqrt{\kappa}}e^{\pm
i\beta(\tilde{\sigma})}.$$ This is the asymptotic wave function in the regions $\beta\rightarrow -\infty$, while around $\beta\sim 0$; i.e. $\tilde{\sigma}\sim\sqrt{\kappa}$, $f^i_\ell\sim 2^{-\frac{1}{4}}$. The variation of this fluctuation in terms of small $\sigma$ and the electric field is well shown in Fig.3(a) by considering the real part of the function. The variation of $f^i_\ell$ in terms of $\sigma$ has positive values and goes up as $\sigma$ goes up for all $E$ in general. The influence of $E$ on $f^i_\ell$ appears at $\sigma\approx0.2$.

Now, if $\sigma$ is too large the equation of motion (98) becomes
\beq \Big[ -\partial_{\sigma}^2 +\frac{\ell(\ell+1)}{(1+\lambda E)\sigma^2}\Big] f^i_\ell =  \frac{1-\lambda E}{1+\lambda E} w^2  f^i_\ell. \eeq

The fluctuation solution of this equation is \beq\bra{ll}f^i_\ell(\sigma)&=\alpha
\sqrt{\sigma}BesselJ\Big(\frac{1}{2}\sqrt{1+4\frac{\ell(\ell+1)}{1+\lambda E}},w\sigma\sqrt{\frac{1-\lambda E}{1+\lambda E}}\Big)\\&+\beta\sqrt{\sigma}BesselY\Big(\frac{1}{2}\sqrt{1+4\frac{\ell(\ell+1)}{1+\lambda E}},w\sigma\sqrt{\frac{1-\lambda E}{1+\lambda E}}\Big),\era\eeq with $\alpha$, $\beta$ are constants. The variation of this fluctuation in terms of large $\sigma$ and $E$ is given by Fig.3(b). The values of $f^i_\ell$ are negative and they are going down as $E$ going up.
%**************************
\begin{center}
\includegraphics[width=6in,height=4in]{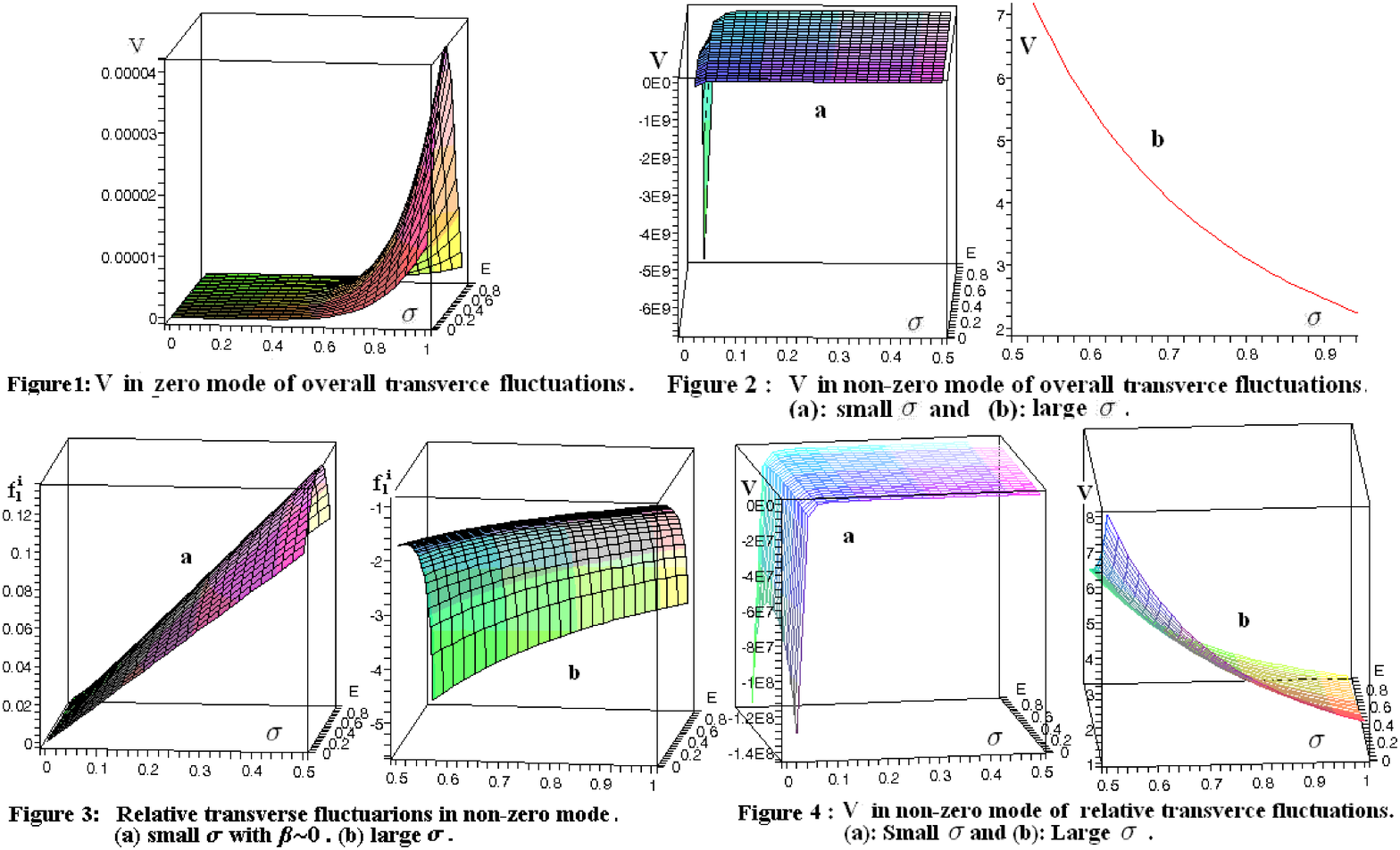}
\end{center}
%**************************
By dealing with the fluctuations (106) (Fig.3(a)) and (108) (Fig.3(b)) in small and large $\sigma$ regions respectively, we remark that is clear that we get different fluctuations from small to large $\sigma$ with a singularity at some stage of $\sigma$ and consequently the system is separated into two regions depending on the electric field.

The potential associated to (108) is $$V(\sigma)=\frac{\ell(\ell+1)}{(1+\lambda E)\sigma^2}.$$ Accordingly to Fig.4(b) describing the variation of $V$, we remark that, $V$ goes down as $\sigma$ goes up and more down as $E$ goes up; i.e. the potential becomes more small as the electric field appears between zero and $E\sim \frac{1}{\lambda}$. The potentials represented by the two figures Fig.4(a) and Fig.4(b) don't have an intersecting point at some stage of $\sigma$. This leads the system to get a singularity which supports the idea that the system is separated into two regions in non-zero modes of relative transverse fluctuations.

Consequently, the D1$\bot$D3 system has Neumann boundary conditions and this is more clear at the presence of electric field. This is proved through this section by discussing different modes and different directions of the fluctuations of the funnel solution and their associated potentials. In the figures representing these variations we set $w$, $\ell$ equals to the unit of $\lambda=1$ in all the treated equations and $N=10$.

\subsection{Overall Transverse Fluctuations in D1$\bot$D5 System}
\hspace{.3in}We extend this study to discuss the electrified fluctuations in D1$\bot$D5 system. we give the equations of motion of the fluctuations and their solutions. Then, we discuss the variation of the potential and the fluctuations in terms of electric field and the spatial coordinate.

We start by considering overall transverse fluctuations in zero mode. This type of fluctuations is given as $\delta \phi^m (\sigma,t)=f^m (\sigma,t)I_N$, $m=6,7,8$. Plugging this fluctuation into the full ($N,N_f$)-string action (15'), together with the funnel (16) the action is found to be \beq S=-NT_1 \int d^2\sigma \Big((1+\lambda E)A -\frac{1}{2}(1-\lambda^2 E^2 )\lambda^2 A(\dot{f}^m)^2+\frac{1}{2}(1+\lambda E)\lambda^2 (\partial_\sigma f^m)^2 +...\Big). \eeq where \beq A=(1+\frac{4R(\sigma)^4}{c\lambda^2 })^2 ,\eeq with the quadratic terms in $f^m$ were the only terms retained in the action. The linearized equation of motion of the fluctuation is then \beq\Big[ (1-\lambda E)\Big( 1+\frac{4R(\sigma)^4}{c\lambda^2 }\Big)^2 \partial^{2}_{t}-\partial^{2}_{\sigma}\Big] f^m =0.\eeq We consider small $R$ which is given by (17) in the second section. We insert its expression in the last equation and the equation of motion becomes \beq\Big[ (1-\lambda E)\Big( 1+\frac{n^2 \lambda^2 }{16(1-\lambda^2E^2)^2\sigma^4}\Big)^2 \partial^{2}_{t}-\partial^{2}_{\sigma}\Big] f^m =0\eeq for large
$n$.

Let's consider the fluctuation in the following form \beq
f^m =\phi(\sigma)e^{-iwt}\delta x^m, \eeq with $\delta x^m$, $m=6,7,8$, the
direction of the fluctuation. The equation (112) becomes
\beq\Big[-\partial^{2}_{\sigma} -w^2 (1-\lambda E)(\frac{n^2
\lambda^2 }{8(1-\lambda^2E^2)^2\sigma^4}+\frac{n^4 \lambda^4 }{16^2(1-\lambda^2E^2)^4\sigma^8}) \Big]
\phi =w^2 (1-\lambda E)\phi.\eeq The potential  of this system is \beq V= -w^2 (1-\lambda E)(\frac{n^2
\lambda^2 }{8(1-\lambda^2E^2)^2\sigma^4}+\frac{n^4 \lambda^4 }{16^2(1-\lambda^2E^2)^4\sigma^8})\eeq depending on the electric field $E$ with $E\in [0,\frac{1}{\lambda}[$.
\subsubsection{Small $\sigma$ Limit}
The equation (114) is complicated and to simplify the calculations we start by considering the small $\sigma$ and $\frac{1}{\sigma^8}$ dominates in (114) and (115). We then discuss the equation
\beq\Big[-\partial^{2}_{\sigma} -w^2 (1-\lambda E)\frac{n^4 \lambda^4 }{16^2(1-\lambda^2E^2)^4\sigma^8}\Big]
\phi =w^2 (1-\lambda E)\phi,\eeq and the potential is reduced to \beq V= -w^2 (1-\lambda E)\frac{n^4 \lambda^4}{16^2 (1-\lambda^2E^2)^4\sigma^8}.\eeq

As shown in Fig.5(a), the potential $V$ tends to $-\infty$ until some values of $E$ when $E$ and $\sigma$ are close to zero, and once $E$ is close to the inverse of $\lambda$ the potential is zero for all small $\sigma$. We consider in this case $\sigma\in]0,0.5]$ in the unit of $\lambda$ with $\lambda=1$, $w=1$, $n=10$ and $E\in[0,1[$.

To solve the differential equation (116), we consider the total differential on the fluctuation. Let's denote $\partial_{\sigma}\phi\equiv \phi'$. Since $\phi$ depends only on $\sigma$ we find $\frac{d\phi}{d\sigma}=\partial_{\sigma}\phi$. We rewrite the equation (116) in this form \beq \frac{1}{\phi}\frac{d\phi'}{d\sigma}=-w^2 (1-\lambda E)[\frac{n^4 \lambda^4 }{16^2(1-\lambda^2E^2)^4\sigma^8}+1].\eeq An integral formula can be written as follows \beq \int\limits_{0}^{\phi'}\frac{d\phi'}{\phi}=-\int\limits_{0}^{\sigma} w^2 (1-\lambda E)[\frac{n^4 \lambda^4 }{16^2(1-\lambda^2E^2)^4\sigma^8}+1]d\sigma,\eeq which gives \beq \frac{\phi'}{\phi}=-w^2 (1-\lambda E)[-\frac{n^4 \lambda^4 }{16^2(1-\lambda^2E^2)^4\times 7\sigma^7}+\sigma]+\alpha.\eeq We integrate again the following \beq\int\limits_{0}^{\phi}\frac{d\phi}{\phi}=-\int\limits_{0}^{\sigma}(w^2 (1-\lambda E)[-\frac{n^4 \lambda^4 }{16^2\times 7(1-\lambda^2E^2)^4\sigma^7}+\sigma]+\alpha) d\sigma.\eeq We get \beq\ln\phi=-w^2 (1-\lambda E)[-\frac{n^4 \lambda^4 }{16^2\times 42(1-\lambda^2E^2)^4\sigma^6}+\frac{2}{\sigma^2}]+\alpha\sigma+\beta,\eeq and the fluctuation in small $\sigma$ region is found to be \beq \phi (\sigma)=\beta e^{-w^2 (1-\lambda E)[-\frac{n^4 \lambda^4 }{16^2\times 42(1-\lambda^2E^2)^4\sigma^6}+\frac{\sigma^2}{2}]+\alpha\sigma},\eeq with $\beta$ and $\alpha$ are constants.

We plot the progress of the obtained fluctuation in (Fig.6(a)). First we consider the constants $\beta=1=\alpha$, then the small spatial coordinate in the interval $[0,0.5]$ with the unit of $\lambda=1$, $w=1$ and $n=4$. As above the electric field is in $[0,1[$. We see that at the absence of the electric field there is no fluctuations at all and this phenomenon continues for the small values of $E$. When $E\approx0.5$ the fluctuation appears from $\sigma=0.15$ and goes down as $\sigma$ and $E$ go up.

\subsubsection{Large $\sigma$ Limit}
In the large $\sigma$ case, the equation (114) becomes \beq\Big[-\partial^{2}_{\sigma} -w^2 (1-\lambda E)\frac{n^2\lambda^2 }{8(1-\lambda^2E^2)^2\sigma^4}\Big]\phi =w^2 (1-\lambda E)\phi\eeq and the potential is \beq V= -w^2 (1-\lambda E)\frac{n^2\lambda^2 }{8(1-\lambda^2E^2)^2\sigma^4}.\eeq By plotting the progress of this potential (Fig.5(b)) we consider the large spatial coordinate in the interval $[0.5,1]$ and $E\in[0,1[$ in the unit of $\lambda=1$, $w=1$ with $n=10$. The obtained figure shows that $V$ has in general higher values than the ones obtained in small $\sigma$ case (Fig.5(a) describing (117)). Specially, for the first values of $\sigma$, $V$ goes up from negative values to be close to zero for almost values of $E$ until $E$ is close to $\frac{1}{\lambda}$, approximately from $E=0.8$ where $V\approx -0.02$, we remark that $V$ has small variation in [0.8,1] region of $\sigma$. By contrary, in figure 5(a), when $\sigma=0.5$ which is the last value of $\sigma$ in that case we find $V$ is already zero for all $E$. Consequently, these two potentials (117) and (125) show a big gap to go from one system to other that they describe, meaning that our system is separated into two regions; small and large $\sigma$ depending on $E$.

Now, we should solve the equation of motion of the relative transverse fluctuations (124), in the case of large $\sigma$. We start by defining a new coordinate $$\tilde{\sigma}^2= w^2 (1-\lambda E)\sigma^2$$ and (124) becomes \beq\Big( 1+\frac{n^2 \lambda^2 w^4 (1-\lambda E)^2}{8(1-\lambda^2 E^2)^2\tilde{\sigma}^4} +\partial^{2}_{\tilde{\sigma}}\Big) \phi(\tilde{\sigma}) =0,\eeq with the potential is \beq V(\tilde{\sigma})=\frac{\kappa^2}{\tilde{\sigma}^4}, \eeq and
$$\kappa^2=\frac{n^2 \lambda^2 w^4 (1-\lambda E)^2}{8(1-\lambda^2 E^2)^2}.$$ The equation
(126) is a Schr\"odinger equation for an attractive singular
potential $\propto\tilde{\sigma}^{-4}$ and depends on the single
coupling parameter $\kappa$ with constant positive Schr\"odinger
energy. The solution is then known by making the following
coordinate change \beq\chi(\tilde{\sigma})=\int\limits^{\tilde{\sigma}}_{\sqrt{\kappa}}dy\sqrt{1+\frac{\kappa^2}{y^4}}, \eeq and \beq \Phi=(1+\frac{\kappa^2}{\tilde{\sigma}^4})^{-\frac{1}{4}}\tilde{\Phi}.\eeq Thus, the equation (126) becomes \beq\Big(-\partial^{2}_{\chi}+V(\chi)\Big) \tilde{\Phi}=0,\eeq with \beq V(\chi)=\frac{5\kappa^2}{(\tilde{\sigma}^2
+\frac{\kappa^2}{\tilde{\sigma}^2})^3}.\eeq Then, the fluctuation is found to be \beq
\Phi=(1+\frac{\kappa^2}{\tilde{\sigma}^4})^{-\frac{1}{4}}e^{\pm
i\chi(\tilde{\sigma})}. \eeq This fluctuation has the following
limit; since we are in large $\sigma$ region $\Phi\sim e^{\pm i\chi(\tilde{\sigma})}$. This is the asymptotic wave function in the regions $\chi\rightarrow +\infty$, while around $\chi\sim 0$;
i.e. $\tilde{\sigma}\sim\sqrt{\kappa}$, $\Phi\sim 2^{-\frac{1}{4}}$. Owing to the plotting of the progress of this fluctuation given by Fig.6(b), by considering the real part of the function, we remark that $\Phi$ goes down fast as $E$ goes up for all $\sigma$. When $\sigma=0.5$ the fluctuation gets different values for all $E$ compared to the values gotten in small $\sigma$ region by (123) (Fig.6(a)). These tow figures show that the fluctuations of fuzzy funnel of D1$\bot$D5 branes have a singularity at some stage of $\sigma$ separating the system into two regions; small and large $\sigma$.
%**************************
\begin{center}
\includegraphics[width=4in,height=4in]{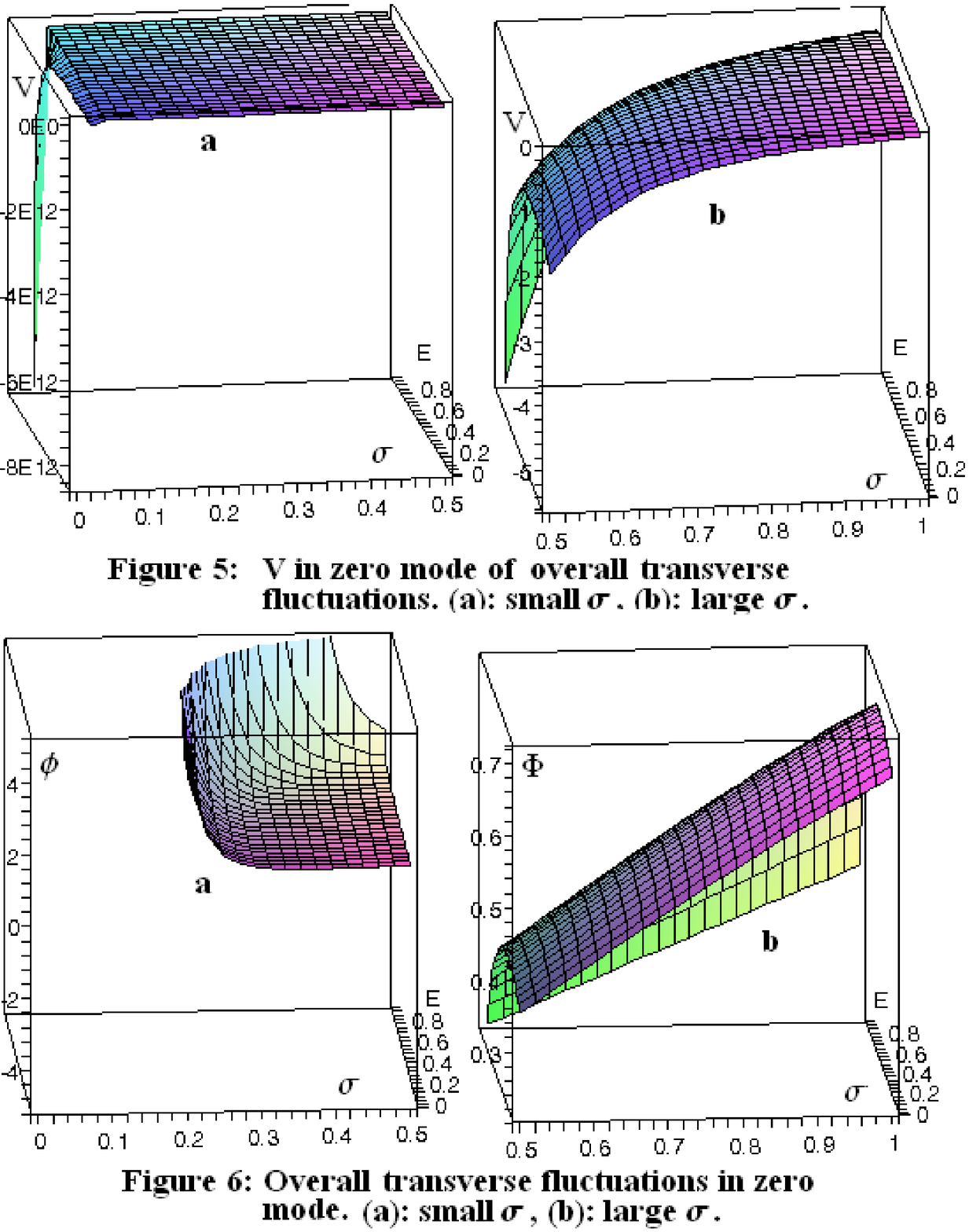}
\end{center}
%**************************

\section{Duality in Electrified D1$\bot$D3 and D1$\bot$D5 Systems}
\subsection{Duality in D1$\bot$D3 Branes}
\hspace{.3in}Now, we see if D1 and D3 descriptions \cite{Gib} in D1$\bot$D3 system discussed in section 2 match or not. As we showed in \cite{brokdual}, the two descriptions D1 and D3 don't have a complete agreement in the presence of a world volume electric field since the energies don't match as we will see in the following, even if their profiles match very well.

The energy is easily derived from the action (8) for the static solution (10). The minimum energy condition is $$\frac{d\phi_i}{d\sigma}=\pm\frac{i}{2}\epsilon_{ijk}[\phi_j,\phi_k],$$ which can be identified as the Nahm equations \cite{nahm}. We insert the ansatz (9) and this implies \beq\hat{R}'=\pm 2\sqrt{1-\lambda^2 E^2}\hat{R}^2.\eeq Using this condition and evaluating the Hamiltonian, $\int d\sigma (DE-L)$, for the dyonic funnel solutions ($L$ is the Lagrangian), the energy is expressed as
$$E_1=T_1 \int d\sigma STr \Big[\frac{\lambda^2 E^2}{\sqrt{1-\lambda^2 E^2}}+\sqrt{1-\lambda^2 E^2}|1+4\lambda^2 \al_j \al_j \hat{R}^4 |\Big].$$ We can manipulate this result by introducing the physical radius $R=\lambda\sqrt{C}|\hat{R}|$ and using $T_1=4\pi^2\ell_s T_3$. It's also useful to $$D
\equiv \frac{1}{N_m}\frac{\de S}{\de E}=\frac{\lambda^2 T_1 E}{\sqrt{1-\lambda^2 E^2}}=\frac{N_e}{N_m}.$$ consider the electric displacement $D$ 
Consequently, the energy from the D1 brane theory is found to be \beq E_1=T_1 \int d\sigma\sqrt{N_m^2
+g_s N_e^2}+T_3 (1-\frac{1}{N_m^2})^{\frac{-1}{2}}\int dR 4\pi R^2 ,
\eeq 
with $g_s$ is the string coupling with $T_1=(\lambda g_s)^{-1}$. The first term comes from collecting the contributions independent of $\hat{R}$. The second term gotten from the terms containing $\hat{R}$ and is used to put these in the form $\hat{R}^2|\hat{R}'|$. Then we have repeatedly applied (133) in producing the second term.

If we consider large $N_m$ the energy is reduced to the
following \beq E_1=T_1 N_m \int d\sigma, \eeq which can be rewritten
in terms of physical radius $R$ as \beq E_1=T_3 N_m \int 4\pi R^2 dR,
\eeq with $T_3 =\frac{T_1}{4\pi^2 \ell_s^2}$. In D3-brane
description the energy (3) becomes \beq E_3=T_3 \int
d^3\sigma\sqrt{1+\lambda^4 \frac{N_m^2 [(N_m +N_e )^2 +N_e^2]}{16r^8
} + 2\lambda^2 \frac{N_m (N_m +N_e )}{4r^4 }+2\lambda^2 \frac{ N_e^2
}{4r^4 }},\eeq such that the magnetic and the electric fields are
given by \beq \stackrel{\rightarrow}{B}=\frac{N_m
}{2r^2}\stackrel{\rightarrow}{r},
\phantom{~~~~}\stackrel{\rightarrow}{E}=\frac{N_e
}{2r^2}\stackrel{\rightarrow}{r}.\eeq In the large $N_m$ limit and
fixed $N_e$, the energy (137) of the spherically symmetric BPS
configuration is reduced to \beq E_3=T_3 \frac{N_m \sqrt{(N_m +N_e
)^2 +N_e^2}}{N_m +N_e} \int 4\pi r^2 dr. \eeq Again we consider large
$N_m$ limit and fixed $N_e$ and we get
$$\frac{ \sqrt{(N_m +N_e )^2 +N_e^2}}{N_m +N_e}\longrightarrow 1.$$
Consequently, for fixed $N_e$ and large $N_m$ limit we have
agreement from both sides (D1 and D3 descriptions) and the energy is
\beq E_3=T_3 N_m \int 4\pi r^2 dr, \eeq in which we identify the
physical radius $R$ from D1 description and $r$ from D3 description.

Now, if we take large $N_m$ limit keeping $N_e/N_m=K$ fixed at any
arbitrary $K > 0$ the result will be different. Thus, from D1
description the energy becomes \beq E_1=T_3 N_m \sqrt{1+g_s K^2}
\int 4\pi R^2 dR, \eeq and from D3 description the energy is \beq
E_3=T_3 \frac{N_m \sqrt{(1+K)^2 +K^2}}{1+K} \int 4\pi R^2 dR . \eeq
Then we have disagreement. Consequently, the presence of electric
field spoils the duality between D1 and D3 descriptions of
intersecting D1-D3 branes.
\subsection{Duality in D1$\bot$D5 Branes}
\hspace{.3in}Although D1$\bot$D5 branes system \cite{cm,f1} is not supersymmetric, the fuzzy funnel
configuration in which the D-strings expand into orthogonal
D5-branes shares many common features with the D3-brane funnel.
Thus, we are interested in establishing whether a similar result
holds also in the case of D1$\bot$D5 branes meaning the presence of
a world volume electric field leads to broken duality or not.

From the D1 description the system is described by the action (15''). We consider static configurations involving five (rather than three) nontrivial scalars, $\phi_i$ with $i=1,...,5$ with the proposed ansatz (16). 

The electric field is fixed by the quantization condition on the
displacement field, $D=\frac{N_f}{N}$, where \beq
D=\frac{1}{N}\frac{\delta S}{\delta E}=\frac{\lambda^2 T_1
E}{\sqrt{1-\lambda^2 E^2}} \eeq after using the equations of motion,
the energy ($\tilde{E}_1 =\int d\sigma (DE-L)$) of the system is
evaluated to be \beq \tilde{E}_1 =\sqrt{N^2 +g_s^2 N_f^2}T_1 \int
d\sigma +\frac{6N}{c}T_5 \int \Omega_4 R^4 dR +NT_1 \int dR +\Delta
E, \eeq with $T_5 =\frac{T_1}{(2\pi \ell_s)^4}$ and the first and
the second terms correspond to the energies of N semi-infinite
strings stretching from $\sigma = 0$ to infinity and of $6N/c$
D5-branes respectively. The contribution of the last terms to
the energy indicates that the configuration is not supersymmetric.
The last contribution is a finite binding energy
$\Delta E =  1.0102Nc^{1/4}T_1 \ell_s$.

The energy of the D1-D5 system from D5 description (subsection 2.2) is evaluated to be \beq \tilde{E}_5 =\frac{NT_1 }{\sqrt{1-\Big(\frac{g_s N_f}{\sqrt{N^2 +g_s^2 N_f^2}}\Big)^2 }}\int d\sigma
+\frac{6N}{c}T_5 \int \Omega_4 R^4 dR +NT_1 \int dR +\Delta E,\eeq
with $\Delta E$ is the same one found above from D1 description. In the absence of electric field, it's clear that by identifying the profiles of D1 and D5 descriptions in the limit of $N$ we could get complete agreement for the geometry and the energy determined by the two dual approaches. Now, in the presence of an electric field it seems there is also agreement. We compare the energy from the D1 description (144) and the energy from the D5 description (145). If we consider the large $N$ limit and fixed $N_f$ the first term of $\tilde{E}_1$ becomes \beq \sqrt{N^2+g_s^2 N_f^2}T_1 \int d\sigma \longrightarrow NT_1 \int d\sigma ,\eeq and the first term of $\tilde{E}_5$ goes to the following value \beq\frac{NT_1 }{\sqrt{1-\Big( \frac{g_s N_f}{\sqrt{N^2 +g_s^2
N_f^2}}\Big)^2 }}\int d\sigma \longrightarrow NT_1 \int d\sigma ,\eeq which proves the agreement at large $N$.

Now, let's fix the value $\frac{g_s N_f}{N}$ to be one value $M$ which can't be neglected at large limit of $N$. Thus, if $N$ is large the last two limits (146) and (147) become \beq \sqrt{N^2 +g_s^2 N_f^2}T_1
\int d\sigma \longrightarrow NT_1 \sqrt{1+M^2}\int d\sigma \eeq and \beq \frac{NT_1 }{\sqrt{1-\Big( \frac{g_s N_f}{\sqrt{N^2 +g_s^2 N_f^2}}\Big)^2 }}\int d\sigma \longrightarrow \frac{NT_1}{\sqrt{1-\frac{M^2}{1+M^2}}}\int d\sigma.\eeq The right hand term of (149) is equal to the right hand term of (148) $$\frac{NT_1}{\frac{1}{\sqrt{1+M^2}}\sqrt{1+M^2-M^2}}=NT_1 \sqrt{1+M^2}.$$ Then, this implies agreement of the two duals at the level of energy of the two descriptions. Consequently, the duality in D1$\bot$D5 branes is unbroken by switching on the electric field.

\section{D2-Brane and Generalized Maxwell Theory}
\hspace{.3in}One of the field-theories describing anyons is the model where the matter is interacting with the Chern-Simons (CS) gauge field \cite{GCS}. In reference \cite{GConn}, Stern has introduced another approach to treat anyons that does not require the CS term, but introduces a generalized connection to which the conserved U(1) current is coupled in a gauge invariant way \cite{Any}. In this model the gauge field is dynamical and the potential has the confining nature which makes the model different \cite{GC}.

In this section, we treat the same system but on the two-sphere.
Among the main results in this work is the change of the potential's
nature; there is no confinement any more, and the
disappearance of the confinement in the two-sphere case for the exotic
system is very interesting result. It was shown in \cite{conf} that
compact Maxwell theory in (2+1)-dimensions confines permanently
electric test charges and the usual two-dimensional Coulomb
potential is $V (R) \sim lnR$. Since the electrostatic potential has
the form $V (R)\sim R$ and holds for all values of the gauge
coupling, the compact (2+1)-dimensional Maxwell theory does not
exhibit any phase transition, i.e., the confinement is permanent. In
the present paper, things are changed by treating the exotic
system in higher dimensions and $V(R)\sim \frac{1}{R}$ with $R$ is the
distance between two opposite charged exotic particles. Another
important result we get is at the level of energy; D2-brane gets a
higher energy if the radius $r$ of the two-sphere goes to infinity and
it is higher if the number of charges is large which makes the
system very special.

\subsection{Generalized Connection and Anyons}
\hspace{.3in}The simplest way to realize fractional statistics characterizing anyons in (2+1) dimensional space-time is usually by adding a Chern-Simons term to the action. Recently, a novel way was introduced in \cite{GConn} to describe anyons without a Chern-Simons term. Thus, a generalized connection was considered in (2+1)-dimensions denoted $A_\mu^\te$, $\mu=0,1,2$. The gauge theory is defined by the following Lagrangian \beq L_\te=-\frac{1}{4}F_{\mu\nu}F^{\mu\nu}+J^\mu A_\mu^\te \eeq with $A_\mu^\te\equiv A_\mu+\frac{\te}{2}\ep_{\mu\nu\rho}F^{\nu\rho}$ and $\te$ is real parameter in Minkowski space. The Lagrangian $L_\te$ describes Maxwell theory that couples to the current via the generalized connection rather than the usual one. This coupling is gauge invariant as long as $J^{\mu}$ is a conserved external current. In this theory, the gauge fields are dynamical and the canonical momenta are $\pi^\mu =F^{\mu 0}+\te\ep^{0\mu\nu}J_\nu$ which results in the usual primary constraint $\pi^0 =0$ and $\pi^i=F^{i 0}+\te\ep^{0ij}J_j$ ($i,j=1,2$). Thus the magnetic field is $B=\ep_{ij}\partial^i A^j$ and the electric field is $E^i =\pi^i-\te\ep^{ij}J_j$.

Now, accordingly to (150), the equations of motion for $A_\mu$ is \beq\partial^\nu \partial_\nu A_\mu =J_\mu+ \te\ep_{\mu\nu\rho }\partial^\nu J^\rho.\eeq Then, we consider the simplest case of a static pointlike particle located at the origin which is described by $J^0=e\de^{(2)}(x)$. By solving (151) for the gauge field one finds $$A_0 =\frac{\ln r}{2\pi}, \phantom{~~~~~}A_1 =\frac{\te x_2}{2\pi r^2}, \phantom{~~~~~}A_2 =\frac{\te x_1}{2\pi r^2},$$ with $r^2=x_1^2 + x_2^2$. This background describes one unit of an electrically charged particle and an infinitely thin magnetic flux with total flux $\te$ both located at the origin and the shift in the statistics of the particle is fixed by the Aharonov-Bohm effect to be \beq \Delta\phi =\te. \eeq with $\phi=\int\limits_{P}A_i dx_i$ the phase acquired by the pointlike particle traveling along some path $P$. $\Delta\phi$ is the phase difference between any two paths at the same endpoints. We note that in the case of Chern-Simons theory, the phase is two times $\te$ and this is due to the fact that the charged particle is winding around a magnetic flux
while in the present theory we also have the contribution of a flux tube winding around the charged particle. Another reason is that with the $A_\mu^\te$ construction a long range electric field is also generated which couples to the current and gives exactly the same phase.

For a static charged particle located at the origin and $J^i =0$,
the static electromagnetic fields are \beq \bra{ll}
B(x)=e\te\de^{(2)}(x)\\
E_i (x)=-\frac{e}{2\pi}\frac{x_i}{r^2} \era \eeq and the total
magnetic flux attached to $N$ charged particles is \beq \Phi=\int_V
d^2 x B(x)=e\te N. \eeq We note that the both $L_{CS}$ (the
lagrangian in Chern-Simons theory) and $L_\te$ lead to fractional
statistics by the same mechanism of attaching a magnetic flux to the
charged particles but the physics they describe is quite different.
We remark, for example, that in this theory the interaction
potential is an object of considerable interest \cite{GC}. The
potential has confining nature; it grows to infinity when the
natural separation of the physical degrees of freedom grow, but in
the Maxwell-Chern-Simons theory, the Chern-Simons term turns the
electric and magnetic fields massive leading to a screening
potential between static charges.
\subsection{Anyons and Fuzzy Two-Sphere}
\hspace{.3in}Now, let us consider exotic particle moving on a
two-sphere instead of a plane in the background of a monopole put at
the origin. First, the two-sphere is $S^2\sim CP^1
=\frac{SU(2)}{U(1)}$ and the representations of $SU(2)$ are given by
the standard angular momentum theory.

The coordinates of fuzzy two-sphere are given by the $SU(2)$ algebra
\beq \bra{rl} \lbrack X_i ,X_j \rbrack =i\al\ep_{ijk}X_k ,&X_i =\al
L_i , \era \eeq $L_i$ is the total angular momentum with the
representation to be the spin $\ell$ and $\al$ is a dimensionful
constant. We note that around the north pole of $S^2$ labeled by
$L_3 =\ell$, the fuzzy two-sphere algebra becomes a noncommutative
plane if $\ell\rightarrow\infty$, \beq \lbrack X_i ,X_j \rbrack
=i\al^2 \ell\ep_{ij}I, \eeq with $I$ is the identity.
\subsubsection{Connection}
\hspace{.3in}To construct the connection which goes to the
generalized connection given above when the radius of fuzzy
two-sphere goes to infinity we use the first Hopf map as known in
the literature which is a map from $S^3$ to $S^2$ and naturally
introduces a $U(1)$ bundle on $S^2$. Then, the two-sphere can be
parameterized by two complex coordinates $u_\al$ such that
$u_\al^\star u_\al =1$ with $u^\star_\al$ is the complex conjugate of $u_\al$. A spatial
coordinate $x_i$ on $S^2$ with radius $r$ is written in terms of
$u_\al$'s as \beq x_i =ru^\dagger \sigma_i u \eeq with $\sigma_i$ are
Pauli matrices. The vector potential on $S^2$ is \beq A_i dx_i
=-i\ga u_\al^\star du_\al , \eeq with $\ga$ is integer due to the
Dirac quantization rule.

Thus, the Hopf spinor satisfying (157) is given by \beq u=\Big(
\bra{ll}
u_1\\
u_2 \era\Big)=\frac{1}{\sqrt{2r(r+x_3)}}\Big( \bra{ll}
r+x_3\\
x_1+ix_2 \era\Big)e^{i\chi}, \eeq $e^{i\chi}$ is a $U(1)$ phase. The
connection is defined as \beq A_i dx_i =-i\frac{\hbar}{e}
u_\al^\star du_\al =\frac{\hbar}{2er(r+x_3)}\ep_{ij3}x_j dx_i . \eeq
By considering the motion of an exotic particle (charged
particle-magnetic flux composite) on two-sphere, the monopole charge
is $\frac{\hbar}{2e}=\frac{\te}{4\pi}$ which is identified with the
connection in two dimensional space for $r\rightarrow\infty$
discussed in section 2. By generalizing the spinor to
$(2S+1)$-components spinor $u_{(S)}$, the monopole charge becomes
$\frac{\te}{4\pi}=\frac{\hbar S}{e}$ and $$x_i=\frac{1}{S}ru_{(S)}^\dagger \sigma_{(S)_i} u_{(S)},$$ $x_i x_i
=r^2$, where $\sigma_{(S)_i}$ is the spin $S$ representation of
$SU(2)$.

Now, for simplicity we consider a static particle at $\bf{x'}$. The magnetic and electric fields are given in (135) and the charge-magnetic dipole is defined by the current \beq\bra{lr} J_0=e\delta^{(3)}(\bf{x}-\bf{x'})& J_i
=\frac{\mu_m}{e}\ep_{im}\partial_m J_0 , \era\eeq $\mu_m$ is the dipole's moment.

\subsubsection{Generalized Maxwell Theory}
\hspace{.3in}The Hamiltonian of this system is written as follows
\beq H=\frac{1}{2mr^2 }M_i M_i +\int d^3 x
(-\frac{1}{2}F_{i0}F^{i0}+\frac{1}{4}F_{ij}F^{ij}-\frac{\te}{2}\ep_{ij}J^{0}F^{ij})
\eeq such that for a static point like particle $J_i =0$ and the
primary constraint is $$\pi^0 =0$$ which leads to the secondary
constraint $$\partial_i \pi^i -J^0 =0$$ with $\pi^\mu$ is the
canonical momentum of gauge field $A^\mu$. $M_i$ is the orbital
angular momentum of the charged particle \beq\bra{ll}
M_i &=\ep_{ijk}x_j (-i\hbar\partial_k +eA^\te _k)\\ \\
&=\ep_{ijk}x_j (-i\hbar\partial_k +eA _k
+\frac{e\te}{2}\ep_{knm}F^{nm}+\frac{e\te}{2}\ep_{kn0}F^{n0})
\era\eeq where $i,j,k,n,m=1,2,3$ and $A^\te _k$ is the generalized
connection. The strength field $F^{\mu\nu}$ is \beq\bra{ll}
F^{nm}&=-\frac{\te}{4\pi}\ep_{nml}\frac{x_l}{r^3}\\ \\
F^{n0}&=\frac{\te}{4\pi r(r+x_3)} \ep_{nl3}(\dot{x_l}-\dot{r}x_l
\frac{2r+x_3}{r(r+x_3)}), \era\eeq we note that $\ep_{kn0}F^{n0}=0$
since $\ep_{nl3}\ep_{kn0}=0$ because $l\ne 0$. Then \beq M_i
=\ep_{ijk}x_j (-i\hbar\partial_k +eA_k
-\frac{e\te^2}{4\pi}\frac{x_k}{r^3}). \eeq Thus the Hamiltonian (162)
of this system is reduced to \beq H=\frac{1}{2mr^2 }M_i M_i
+\frac{1}{2}\int d^3 x (E_i ^2 +B^2 ), \eeq with $\ep_{oij}J^0
F^{ij} = \frac{-\te}{4\pi}\ep_{oij}\ep^{ijk}J^0 \frac{x_k}{r^3}
=\frac{-\te}{2\pi}\delta_{0k}J^0\frac{x_k}{r^3} =0$ in (13) since
$k=1,2,3$. Accordingly to (153,154), we calculate the second term
of $H$ in three-dimensional space and the Hamiltonian is \beq H =H_0
+e^2 \te^2 N +\frac{e^2 r}{3\pi},\eeq with \beq H_0 =\frac{1}{2mr^2
}M_i M_i .\eeq The conclusion we get from this subsection is that the
Hamiltonian is different from the one describing the QHE and they are
identified ($H\sim H_0$) only if $N,r\longrightarrow 0$.

\subsubsection{Realization of Fuzzy Two-Sphere}
\hspace{.3in}First, we remark that the orbital angular momentum of
the particle $M_i $ given by (165) satisfy the following deformed
commutation relations \beq \lbrack M_i , M_j \rbrack=
i\hbar\ep_{ijk}(M_k +\frac{e\te}{2\pi r}x_k ). \eeq This means that
the total angular momentum generalizing the $SU(2)$ algebra should
be defined as \beq L_i =M_i -\frac{e\te}{2\pi r}x_i \eeq and we get
\beq\bra{lll}
\lbrack L_i , L_j \rbrack = i\hbar\ep_{ijk}L_k \\ \\
\lbrack L_i , M_j \rbrack = i\hbar\ep_{ijk}M_k \\ \\
\lbrack L_i , x_j \rbrack = i\hbar\ep_{ijk}x_k . \era\eeq
Consequently, by simple calculation we find that
$$\lbrack L_i , H\rbrack =0,$$ then $SU(2)$ symmetry is generated
by $L_i$. We also see that \beq\bra{lll}
M_i M_i &=L_i L_i - (\frac{e\te}{2\pi})^2 \\\\
&=\hbar^2 (l(l+1)-4S^2), \era\eeq with $\ell$ is the eigenvalue of
$L_3$. We set, in what follows, $\ell=n+2S$ ($n=0,1,2...$) with $n$ plays the role of the level index which could be identified with the Landau level index in the Chern-Simons theory and $n=0$ corresponds to the lowest level.

The noncommutative Geometry known as fuzzy two-sphere is described
by the coordinates $X_i$ defined as \beq X_i =\frac{2\pi r}{e\te}L_i , \eeq
and they are related to the commutative coordinates $x_i$ by \beq X_i
=\frac{2\pi r}{e\te}M_i -x_i . \eeq They satisfy the following
commutation relations \beq \lbrack X_i , X_j \rbrack
=i\hbar\ep_{ijk}\frac{2\pi r}{e\te}X_k , \eeq and the fuzzy
two-sphere is realized for the motion of exotic particle on a
two-sphere. Its radius is given by the quadratic Casimir of $SU(2)$
\beq r^2 =\hbar^2 (\frac{2\pi r}{e\te})^2 \ell(\ell+1). \eeq

The exotic particles obey the cyclotron motion as well-known in the planar system, in which the magnetic field causes them to follow a circular path or cyclotron orbit. According to (172,173), we get the radius of the cyclotron motion $r^c _n$ in the $n$-th level \beq r^c _n =\frac{2\pi r}{e\te}\hbar\sqrt{2S(2n+1)+n(n+1)},\eeq which is given by $(r^c _n)^2 =(\frac{2\pi r}{e\te})^2M_iM_i$. Owing to (174), this radius is related to the coordinates $X_i$ and $x_i$ and it is given by $(r^c _n)^2 =(X_i+x_i)(X_i+x_i)$.

For the lowest level we get \beq r^c _0 =\frac{r}{\sqrt{2S}},\eeq which is identified with the one obtained in the lowest Landau Level discussed in \cite{fuzzyQHE}. Also we remark that $r^c_0$ is much smaller than $r$ in the lowest level $n=0$ and in a strong magnetic field limit, and $x_i$ are identified with $X_i$.

\subsubsection{Energy}
\hspace{.3in}Owing to (172), the energy eigenvalue of $H$ (167) is
\beq E_n =\frac{\hbar^2 }{2mr^2 }(2S(2n+1)+n(n+1))+e^2 \te^2 N
+\frac{e^2 r}{3\pi}. \eeq Then we notice that this model could be
identified with the one treated in references \cite{fuzzyQHE}
only in the following case: If both of the radius of fuzzy
two-sphere and the number of charges $N$ are vanishingly small; i.e.
$r,N\longrightarrow 0$. Also, we note that $n$ in (179) indicates the
level index which could be identified with the Landau level index in
the Chern-Simons theory only for small $r$ and $N$. The variance
energy between the lowest level $n=0$ and the first level is
$$\Delta E=\frac{\hbar^2 (2S+1)}{mr^2}.$$

As a remark, the lowest level in our system may be
identified with the Lowest Landau level phenomena if the number of
charges is small and $r\longrightarrow 0$ with $\frac{S}{mr}\gg
1$ and the energy induced by the dynamical gauge field is ignored.
Otherwise, if the above case is not satisfied; i.e. $r\gg 1$ or
$N\gg 1$, the model is now totally different. Thus the variance
energy of the system is $$\Delta E=0,$$ and the energy is dominated
by one of the two last terms of (177) or both. Then the energy is
\beq E_n =e^2 \te^2 N +\frac{e^2 r}{3\pi},\phantom{~~~~}\forall n.
\eeq We notice here that the variance energy of the system will
depend only on the variance of the number of particles or the
radius.

Consequently, dealing with the case of exotic particles system in
which we introduce a generalized connection put at the origin of
two-sphere we get a noncommutative geometry. The energy obtained in
this model is very special and very different from the one obtained
in QHE case since the gauge field in this system is dynamical. We
notice that the energy of the gauge field dominates when the radius of
the fuzzy two-sphere goes to infinity; i.e. the flat D2-brane which is a
dual of the fuzzy two-sphere has higher energy. We remark that this result
is definitely different from the one could be obtained in the case
of QHE. In this latter case if $r\longrightarrow\infty$ the fuzzy
two-sphere goes to flat D2-brane having an energy which goes to
zero in this limit.
\subsubsection{Potential}
\hspace{.3in}We complete this section by making another interesting
remark. As known, the potential has a confining nature in
two-dimensional space when the generalized connection is introduced
instead of adding CS-term; i.e., the potential grows to infinity
when the natural separation of the physical degrees of freedom
grows.

After giving the energy we may now proceed to discuss the
interaction energy between pointlike sources in the model under
consideration. This can be done by computing the expectation value
of the energy operator H in a physical state $|\Om\rangle$ by
following the mechanism used in \cite{pot}. We consider the stringy
gauge-invariant $|\bar{\Psi}(y) \Psi(r)\rangle$ state, \beq
|\Om\rangle\equiv |\bar{\Psi}(y) \Psi(y')\rangle
=|\bar{\psi}(y)e^{-ie\int\limits_{y}^{y'}dz^i A_i (z)}
\psi(y')|0\rangle, \eeq where $| 0\rangle$ is the physical
vacuum state and the integral is over the linear spacelike
path starting at $y$ and ending at $y'$, on a fixed time slice. Note
that the strings between exotic particles have been introduced to
have a gauge-invariant state $|\Om\rangle$, in other terms, this
means that the elementary particles (bosons or fermions) are now
dressed by a cloud of gauge fields.

From the foregoing Hamiltonian discussion, we first note that \beq
\pi_i |\bar{\Psi}(y) \Psi(y')\rangle=\bar{\Psi}(y) \Psi(y') \pi_i
| 0\rangle + e\int\limits_{y}^{y'} dz_i \delta^3
(x-z)|\bar{\Psi}(y) \Psi(y')\rangle.\eeq Owing to (166,180) and
the fact that we consider a static pointlike particle; so $\pi_i
=F_{0i}=E_i$, we get the expectation value of the Hamiltonian as
\beq \langle\Om| H|\Om\rangle=\langle 0| H| 0\rangle
+\frac{e^2}{2}\int d^3 x \Big( \int\limits_{y}^{y'} dz_i \delta^3
(x-z)\Big)^2 , \eeq with $x$ and $z$ are three-dimensional vectors.
Remembering that the integrals over $z_i$ are zero except on the
contour of integrations.

The last term of (183) is nothing but the Coulomb interaction plus an
infinite self-energy term. In order to carry out this calculation we
write the path as $z = y+\al(y -y')$ where $\al$ is the parameter
describing the contour. By using the spherical coordinates the
integral under square becomes \beq \int\limits_{y}^{y'} dz_i
\delta^3 (x-z)=\frac{y-y'}{|y-y'|^2}\int\limits_{0}^{1} d\al
\frac{1}{\al}\de(|y-x|,\al|y'-y|)\sum\limits_{\ell,m}Y^* _{\ell
m}(\te' ,\phi')Y _{\ell m}(\te ,\phi). \eeq Using the usual
properties for the spherical harmonics and after subtracting the
self-energy term, we obtain the potential as \beq V=-\frac{e^2
}{4\pi} \frac{1}{|y' -y|}.\eeq

This result lets us to draw attention to the fact that with fuzzy
two-sphere the generalized Maxwell theory doesn't have confining
nature any more which was a special property for anyons described by
generalized Maxwell theory in two-dimensional space. Thus the
problem of confinement could be solved by considering the two-sphere
in stead of two-dimensional flat space.

\section{Discussion and Conclusion}
\hspace{.3in}We considered the abelian and non-abelian BI dynamics of the dyonic string, from section 2 until section 6, such that the electric field $E$ has a limited value. The limit of $E$ attains a maximum value \beq E_{max}=T_1=\frac{1}{\lambda}\eeq (for simplicity we dropped $2\pi$ in all the calculations). This limiting value arises because if $E>\frac{1}{\lambda}$ the action ceases to make physical sense \cite{nphys}. The system becomes unstable. Since The string effectively carries electric charges of equal sign at each of its endpoints, as $E$ increases the charges start to repel each other and stretch the string. For $E$ larger than the critical value (186), the string tension $T_1$ can no longer hold the strings together.

The investigation of excited D3-branes through the two cases of absence or presence of an electric field lead to the fact that by exciting 2 and 3 of its transverse directions the brane develops a spike which is interpreted as an attached bundle of a superposition of coordinates of another brane given as a collective coordinate a long which the brane extends away from the D3-brane. In our study of D3-branes, by exciting 2 and 3 transverse directions we found that a magnetic monopole produces a singularity in the D3-branes transverse displacement which can be interpreted as a superposition of coordinates describing D$p$-branes ($p=2,3$) attached to the D3-brane and the same for the dyonic case. We also obtained another important result that the lowest energy in the intersection branes case is obtained at the level of D1$\bot$D3 branes and the energy is higher if we excite more scalar fields and even more so in the presence of an electric field.

Further in the presence of a world volume electric field, space time dependent solutions can be generalised nicely and for $r_0 = 0$ we observe brane collapses with a speed less than that of light. We have also obtained a generalisation of the BPS solution. An automorphism of our solutions relevant for the $r\rightarrow 1/r$ duality has been discussed.

We also showed that certain excitations of D1$\bot$D3 and D1$\bot$D5 systems can be shown to obey Neumann boundary conditions. In this context, by considering $E\in[1,\frac{1}{\lambda}[$ in D1$\bot$D3 and D1$\bot$D5 branes, we treated the fluctuations of the fuzzy funnel solutions and discussed the associated potentials $V$ in terms of the electric field $E$ and the spatial coordinate $\sigma$. We considered the unit of $\lambda$ in all the figures representing the variations. We limited $\sigma$ to be in the interval [0,0.5] for small $\sigma$ and [0.5,1] for large $\sigma$.

Concerning D1$\bot$D3 system, we gave the variation of $V$ in zero mode of overall transverse fluctuations in Fig.1; this figure shows that the system is looking like separated into two regions depending on the electric field. The potential is stable for $E$ varies from 0 until 0.7 and then $V$ goes up quickly as $E$ close to $\frac{1}{\lambda}$. Then we dealt with the general case of the overall transverse fluctuations which is the non-zero modes. In this case, the idea that the system is divided into two regions appears more clear. We gave the figure representing the potential in Fig.2; we see that at $\sigma=0.5$ the potential gets a singularity. We continued to treat the other kind of the fluctuations, it's the relative transverse fluctuations. These fluctuations are represented  by Fig.3 and the associated potentials by Fig.4; we obtain the same remark that both fluctuations and potential have a singularity at $\sigma=0.5$. This supported the idea that the system is separated into two regions.  

We extended our study to the case of higher dimensions. We treated the electrified fluctuations of D1$\bot$D5 branes and we studied only the zero mode of overall transverse fluctuations. We notice that when the electric field is going up and down the potential of the system is changing and the appearance of the singularity is more clear (Fig.5) and we have the same remarks for the fluctuations of fuzzy funnel solutions as well (Fig.6) which cause the division of the system into tow regions depending on small and large $\sigma$ and also on $E$.

Consequently, the end point of the dyonic strings moves on the brane which means we have Neumann boundary conditions in D1$\bot$D3 and D1$\bot$D5 branes. The physical interpretation is that a string attached to the D3 and D5 branes manifests itself as an electric charge, and the waves on the string cause the end point of the string to freely oscillate. Thus, we realize Polchinski's open string Neumann boundary conditions dynamically by considering non-abelian BI action in D1$\bot$D3 and D1$\bot$D5 systems.

The duality between dyonic D1 and D3 descriptions is no longer valid. We have further argued that the D-string description breaks down. This is perhaps not surprising. Indeed, in this limit [13] have argued that the effective tension of the string goes to zero. Thus, excited strings modes will not be very heavy compared to massless string modes and one might question the validity of the Dirac-Born-Infeld action which retains only the massless modes. Our other interest was the fate of the duality of D1$\bot$D5 branes. We found that the duality between the D1 and D5 descriptions is unbroken in the presence of an electric field. Then, the duality in D1-D5 case is still valid.

In type IIA theory, we have used the generalized Maxwell theory on a two-sphere instead of a flat two-dimensional space. This leads to results totally different from those obtained in the case of QHE in higher dimensions \cite{fuzzyQHE}. By considering the exotic particles described by generalized Maxwell theory, the energy produced by the gauge field contributes to the energy of the system (18) depending on the number of charges and the radius of the sphere. We remark that the energy of gauge field dominates when the radius of two-sphere goes to infinity; i.e. the energy of flat D2-brane is generated by the gauge field leading to increasing energy. We also notice that the energy becomes higher if the number of charges is large. Another important remark is that with a fuzzy two-sphere the static potential for two opposite charged exotic particles loses its confining nature without adding the CS-term; i.e. by plunging the generalized Maxwell theory in a higher dimensional theory, the potential has a screening nature.

\end{document}